\documentclass[reqno,11pt]{article}
\usepackage{psfig, amsmath, amstex_new, amsthm}

\theoremstyle{definition}
\newtheorem{example}{Example}

\makeatletter

\def\enum{\ifnum \@enumdepth >3 \@toodeep\else
        \advance\@enumdepth \@ne 
        \edef\@enumctr{enum\romannumeral\the\@enumdepth}\list
        {\csname label\@enumctr\endcsname}
        {\setlength{\topsep}{1mm}
        \setlength{\parsep}{0mm}
        \setlength{\itemsep}{0mm}
        \setlength{\labelsep}{2mm}
        \settowidth{\leftmargin}{M.}
        \addtolength{\leftmargin}{\labelsep}
        \usecounter{\@enumctr}
        \def\makelabel##1{\hss\llap{##1}}}\fi}

\def\itemiz{\ifnum \@itemdepth >3 \@toodeep\else \advance\@itemdepth \@ne
        \edef\@itemitem{labelitem\romannumeral\the\@itemdepth}%
        \list{\csname\@itemitem\endcsname}{
        \setlength{\topsep}{0mm}
        \setlength{\parsep}{0mm}
        \setlength{\parsep}{0mm}
        \setlength{\itemsep}{0mm}
        \setlength{\labelsep}{2mm}
        \settowidth{\leftmargin}{M.}
        \addtolength{\leftmargin}{\labelsep}
        \def\makelabel##1{\hss\llap{##1}}}\fi}

\newenvironment{myquote}
{\begin{list}{}
{\setlength{\labelwidth}{20mm}
\setlength{\labelsep}{1.5mm}
\setlength{\leftmargin}{20mm}
\addtolength{\leftmargin}{\labelsep}
\setlength{\listparindent}{0mm}}
\item
\begin{flushright}\itshape\small}
{\end{flushright}\end{list}\vspace{5mm}}

\def\captionheadfont@{\scshape}
\def\captionfont@{\small}
\long\def\@makecaption#1#2{%
  \setbox\@tempboxa\vbox{\color@setgroup
    \advance\hsize-3pc\noindent
    \captionfont@\captionheadfont@#1\@xp\@ifnotempty\@xp
        {\@cdr#2\@nil}{.\captionfont@\upshape\enspace#2}%
    \unskip\kern-3pc\par
    \global\setbox\@ne\lastbox\color@endgroup}%
  \ifhbox\@ne % the normal case
    \setbox\@ne\hbox{\unhbox\@ne\unskip\unskip\unpenalty\unkern}%
  \fi
  \ifdim\wd\@tempboxa=\z@ % this means caption will fit on one line
    \setbox\@ne\hbox to\columnwidth{\hss\kern-3pc\box\@ne\hss}%
  \else % tempboxa contained more than one line
    \setbox\@ne\vbox{\unvbox\@tempboxa\parskip\z@skip
        \noindent\unhbox\@ne\advance\hsize-3pc\par}%
\fi
  \ifnum\@tempcnta<64 % if the float IS a figure...
    \addvspace\abovecaptionskip
    \moveright 1.5pc\box\@ne
  \else % if the float IS NOT a figure...
    \moveright 1.5pc\box\@ne
    \nobreak
    \vskip\belowcaptionskip
  \fi
\relax
}

\makeatother

\def\math#1{\ifmmode
\mathchoice{\mbox{$\displaystyle\rm#1$}}
{\mbox{$\textstyle\rm#1$}}
{\mbox{$\scriptstyle\rm#1$}}
{\mbox{$\scriptscriptstyle\rm#1$}}\else
{\mbox{$\rm#1$}}\fi}		% roman style with adaptable size

\def\smath#1{\ifmmode
\mathchoice{\mbox{$\textstyle#1$}}
{\mbox{$\scriptstyle#1$}}
{\mbox{$\scriptscriptstyle#1$}}
{\mbox{$\scriptscriptstyle#1$}}\else
{\mbox{$\textstyle#1$}}\fi}	% small size

\def\vec#1{\ifmmode
\mathchoice{\mbox{$\displaystyle\bf#1$}}
{\mbox{$\textstyle\bf#1$}}
{\mbox{$\scriptstyle\bf#1$}}
{\mbox{$\scriptscriptstyle\bf#1$}}\else
{\mbox{$\bf#1$}}\fi}		% vectors in bold face

\DeclareMathSymbol{\leqsymb}{\mathalpha}{AMSa}{"36}
\def\leqs{\;\leqsymb\;}
\DeclareMathSymbol{\geqsymb}{\mathalpha}{AMSa}{"3E}
\def\geqs{\;\geqsymb\;}
\DeclareMathSymbol{\gtreqqlesssymb}{\mathalpha}{AMSa}{"54}

\newcommand{\field}[1]{\mathbb{#1}}

\newcommand{\R}{\field{R}\,}	% real numbers
\newcommand{\cA}{{\mathcal A}}	% calligraphic A
\newcommand{\cC}{{\mathcal C}}	% calligraphic C
\DeclareMathOperator{\e}{e}		% natural number
\DeclareMathOperator{\thyp}{th}		% hyperbolic tangent
\DeclareMathOperator{\dd}{d}		% integration measure
\DeclareMathOperator{\defby}{\raisebox{0.35pt}{\math{:}}\!\!=}
\DeclareMathOperator{\bydef}{=\!\!\raisebox{0.35pt}{\math{:}}}

\def\diff#1{\cC^{#1}}	% differentiability
\def\dx#1{\dd\!#1}		% integration measure
\def\dpar#1#2{\frac{\partial #1}{\partial #2}}	% partial derivative
\def\sdpar#1#2{\partial_{#2}#1}		% small partial derivative
\def\dtot#1#2{\frac{\dx{#1}}{\dx{#2}}}	% total derivative
\def\abs#1{\lvert#1\rvert}		% absolute value
\def\biggbrak#1{\biggl[#1\biggr]}	% bigg bracket
\def\O#1{{\math O}(#1)}		% orthogonal matrices
\def\Order#1{{\mathcal O}(#1)}	% Order relation
\def\fix#1{#1^{\star}}		% fixed point
\def\nth#1{\ensuremath{#1^{\math{th}}}}	% nth 
\def\defwd#1{{\bf#1}}		% defined word

\def\ie{i.e.,\ }

\def\Liap{Lyapunov}
\def\DynSys{Dynamical Systems}
\def\DynSy{dynamical system}

\def\nbh{neighborhood}
\def\monoton{monotonous}
\def\dss{\displaystyle}

\def\eps{\varepsilon}

\def\x{\xi}

\def\sub#1{_{\smath{#1}}}	% small subscript
\def\ez{\eps\sub{0}}

\DeclareMathSymbol{\gordsymb}{\mathalpha}{AMSa}{"3C}
\DeclareMathSymbol{\lordsymb}{\mathalpha}{AMSa}{"34}

\def\sord{\approx}		% equivalence relation
\def\ord{\sim}			% equivalence relation
\def\crit#1{#1_{\math{c}}}	% critical value

\def\figref#1{Fig.\ \ref{#1}}	% reference to a figure
\def\tabref#1{Table \ref{#1}}	% reference to a table

\def\writefig#1 #2 #3 {\rlap{\kern #1 truecm
\raise #2 truecm \hbox{\protect{\small #3}}}}
\def\figtext#1{\smash{\hbox{#1}}
\vspace{-5mm}}

\def\bibtitle#1#2{#1, {\em #2}}                         % authors, title
\def\bibref#1#2#3#4#5{#1 {\bf #2}:#3--#4, (#5)}        % review, year
\def\bibarticle#1#2#3#4#5#6#7{\bibtitle{#1}{#2},
\bibref{#3}{#4}{#5}{#6}{#7}.}
\def\bibpreprint#1#2#3#4{#1, {\em #2}, preprint {\tt #3}, (#4).}
\def\bibbook#1#2#3#4{#1, {\em #2}, (#3, #4).}

\def\AJM{Amer.\ J.\ Math.}
\def\AMS{American Mathematical Society}
\def\AP{Ann.\ Physics}
\def\CMP{Comm.\ Math.\ Phys.}
\def\DE{Diff.\ Equ.}
\def\Dokl{Dokl.\ Akad.\ Nauk SSSR}
\def\Doklt{Sov.\ Math.\ Dokl.}
\def\DU{Diff.\ Urav.\ }
\def\JPA{J.\ Phys.\ A}
\def\JPCM{J.\ Phys: Condens.\ Matter}
\def\JSP{J.\ Stat.\ Phys.}
\def\Nat{Nature}
\def\PA{Physica A}
\def\PD{Physica D}
\def\PRA{Phys.\ Rev.\ A}
\def\PRB{Phys.\ Rev.\ B}
\def\PRE{Phys.\ Rev.\ E}
\def\PRL{Phys.\ Rev.\ Letters}
\def\RMP{Rev.\ Mod.\ Phys.}
\def\SAM{Stud.\ in Appl.\ Math.}
\def\SIAM{SIAM J.\ Appl.\ Math.}

\begin{document}

%%%%%%%%%%%%%%%%%%%%%%%%%%%%%%%%%%%%%%%%%%%%%%%%%%%%%%%%%%%%%%%%%%%%%%%%%%%%%%%%%

\title{Hysteresis in Adiabatic Dynamical Systems:\\
an Introduction}

\author{
N. Berglund \\
{\it Institut de Physique Th\'eorique} \\
{\it Ecole Polytechnique F\'ed\'erale de Lausanne} \\
{\it PHB-Ecublens, CH-1015 Lausanne, Switzerland} \\
{\rm e-mail: }{\tt berglund@iptsg.epfl.ch} \\
}

\date{April 16, 1998}

\maketitle

\thispagestyle{empty}

\begin{abstract}
We give a nontechnical description of the behaviour of dynamical systems
governed by two distinct time scales. We discuss in particular memory
effects, such as bifurcation delay and hysteresis, and comment the scaling
behaviour of hysteresis cycles. These properties are illustrated on a few
simple examples.
\end{abstract}

\vspace{30mm}

\vspace{40mm}
\noindent
{\bf Note:}
This text is the introduction to, and summary of the author's Ph.D.\
dissertation. Some references are to chapters of the dissertation.
Postscript files are available at 

\noindent
{\tt http://dpwww.epfl.ch/instituts/ipt/berglund/these.html}

\noindent
Please contact the author for further information.

\newpage
\thispagestyle{empty}
\phantom{empty page}
\setcounter{page}{0}
\newpage

%%%%%%%%%%%%%%%%%%%%%%%%%%%%%%%%%%%%%%%%%%%%%%%%%%%%%%%%%%%%%%%%%%%%%%%%%%%%%%%%%

\begin{myquote}
\iffalse
``For those who like this sort of thing, this is the sort of thing they
like.''

Abraham Lincoln
\fi

``Try not to have a good time\dots This is supposed to be educational.''

Charles Schulz

\end{myquote}

%%%%%%%%%%%%%%%%%%%%%%%%%%%%%%%%%%%%%%%%%%%%%%%%%%%%%%%%%%%%%%%%%%%%%%%%%%%%%%%%%

\section{Introduction}
\label{sec_intech}

%%%%%%%%%%%%%%%%%%%%%%%%%%%%%%%%%%%%%%%%%%%%%%%%%%%%%%%%%%%%%%%%%%%%%%%%%%%%%%%%%

%%%%%%%%%%%%%%%%%%%%%%%%%%%%%%%%%%%%%%%%%%%%%%%%%%%%%%%%%%%%%%%%%%%%%%%%%%%%%%%%%

\subsection{Dynamic Variables and Parameters}
\label{ssec_inpar}

%%%%%%%%%%%%%%%%%%%%%%%%%%%%%%%%%%%%%%%%%%%%%%%%%%%%%%%%%%%%%%%%%%%%%%%%%%%%%%%%%

Since the discovery of Newton's equation and its application to the study
of the Solar System, it has become apparent that an important number of
physical problems could be modeled, more or less accurately, by ordinary
differential equations (ODEs). Sometimes, these equations are direct
consequences of the fundamental laws of Physics, like Newton's equation
(for classical mechanical systems) or Maxwell's equations (for
electromagnetic problems). Macroscopic systems, for which we cannot neglect
the fact that they are composed of a very large number of atoms or
molecules, may sometimes be modeled by somewhat more phenomenological
laws, taking into account the interaction of a small number of effective
degrees of freedom. This applies to the equations of thermodynamics
(applicable for instance to kinetics of chemical reactions), master
equations (lasers) or mean field equations (phase transitions). There also
exist a number of systems, which are not directly related to Physics, but
are nevertheless modeled, on a very phenomenological level, by ODEs: this
is the case, for instance, for population dynamics in ecology. 

\begin{figure}
 \centerline{\psfig{figure=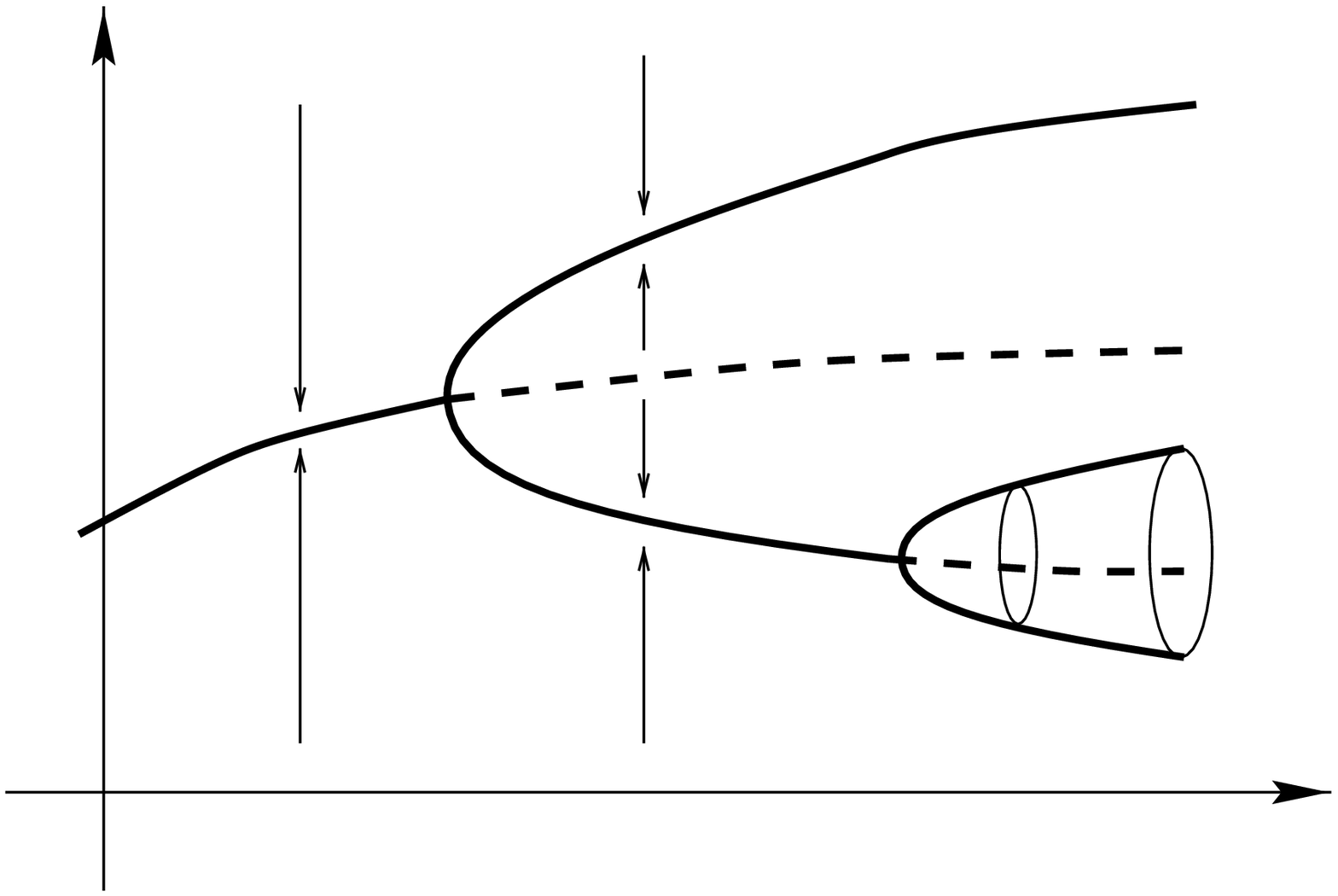,height=50mm,clip=t}}
 \figtext{
 	\writefig	2.3	4.8	\small{asymptotic}
 	\writefig	2.3	4.3	\small{state}
 	\writefig	9.0	0.6	\small{parameter}
 }
% \vspace{2mm}
 \caption[Bifurcation diagram]
 {Example of a bifurcation diagram. For each value of the parameter, one
 plots the asymptotic state of the system. In this example, there is a
 unique stable equilibrium state for small parameter (thick full line). At
 some parameter value, this equilibrium becomes unstable (dotted line),
 while two new stable equilibria are formed. One of them is then replaced
 by a limit cycle, \ie a stable periodic orbit. To determine this diagram
 experimentally, one should fix a value of the parameter and an initial
 condition, and wait for the system to relax to equilibrium (vertical
 arrows). This procedure should be repeated for several parameter values.}
\label{fig_in1}
\end{figure}

\begin{table}[b]
\begin{center}
\begin{tabular}{|l|l|l|}
\hline
{\bf System} 		& {\bf Dynamic variables}& {\bf Parameters} 	\\
\hline	
\hline	
Mechanical system	& Positions and momenta	& External driving force\\
\hline	
Electric device		& Charges and currents	& Power supply, 	\\
			& 			& tunable resistance	\\
\hline	
Chemical reaction	& Concentration of 	& Supply flux, 		\\
			& reacting substances	& temperature 		\\
\hline	
Laser			& Level population,	& External field	\\
			& internal field	& 			\\
\hline	
Magnet			& Order parameter	& Magnetic field,	\\
			& (magnetization)	& temperature		\\
\hline	
Population dynamics	& Number of individuals	& Climate,		\\
			& of each species	& reproduction rate	\\
\hline
\end{tabular}
\end{center}
%\vspace{2mm}
\caption[Dynamic variables and parameters]
{Examples of systems which can be modeled by ODEs, with associated
dynamic variables and parameters.}
\label{tab_in1}
\end{table}
When we consider some specific examples, like those given in
\tabref{tab_in1}, we realize that such differential equations will depend
on two kinds of variables: \defwd{dynamic
variables}\index{dynamic!variable} and \defwd{parameters}\index{parameter}.
As far as the mathematical model is concerned, the distinction between
these two types of variables is clear:
\begin{itemiz}
\item	dynamic variables define the state of the system; their role is
	twofold: on one hand, they evolve in time, specifying the state of
	the system at each instant; on the other hand, they determine the
	future evolution of the system;
\item	parameters also influence the future evolution, but their value
	remains fixed; in fact, a different \DynSy\ is obtained for each
	value of the parameters.
\end{itemiz}
Are the ``parameters'' of \tabref{tab_in1} really always fixed? Let us
examine more closely different kinds of parameters which may appear in a
physical experiment. We may distinguish the following types:
\begin{itemiz}
\item	parameters which are related to physical constants or technical
	specifications of the experimental set--up, and are, therefore,
	fixed during the experiment; this applies to masses and coupling
	constants of particles, and dimensions of a cavity or reactor;
\item	\defwd{control parameters}\index{control parameter}, which can be
	accurately tuned, say by turning a knob of the experimental device;
	this may be the case for the supply voltage of an electric device,
	an applied external field, or the temperature difference between two
	sides of a cavity;
\item	parameters that one would like to maintain fixed, but which are not
	so easy to control in a real experiment, like a supply flux of
	chemicals, or the temperature in a reactor.
\end{itemiz}
One usually characterizes a \DynSy\ by its \defwd{bifurcation
diagram}\index{bifurcation!diagram}\index{diagram!bifurcation},
representing the asymptotic state (which may be stationary, periodic or
more complicated) against the control parameter (\figref{fig_in1}). What do
we mean when we say that the bifurcation diagram is determined
experimentally by varying the control parameter?

According to the mathematical modeling, the bifurcation diagram should
be determined as follows. Fix the control parameter and choose an initial
state for the system. Let the system evolve until it has reached an
asymptotic state. Repeat this procedure for different initial conditions, in
order to find other possible asymptotic states. Then increase the control
parameter, reset the initial state, and repeat the whole experiment. Apply
this procedure for the desired set of parameter values, and plot the
asymptotic state(s) against the control parameter.

In practice, it is not always possible to carry out this rather elaborate
program. We may not have the time to wait for the system relaxing to
equilibrium for each parameter value, or we may not be able to reset the
initial condition. In fact, it is very tempting\footnote{Every person who
has ever seen an experimental device with a knob for the control parameter
knows that it is indeed {\em very} difficult not to turn this knob during
the experiment.} to turn slowly the knob controlling the parameter during
the experiment, in the hope that if this parameter variation is
sufficiently slow, it will not affect the bifurcation diagram very much.

Is this hope justified? The answer to this question is not immediate at all.
It requires a precise understanding of the relation that exists between, on
one hand, a one--parameter family of autonomous \DynSys\ and, on the other
hand, the system with slowly time--dependent parameter. This relation is by
no means trivial in all cases, since memory effects, in particular
hysteresis, may show up in such systems. An understanding of this relation
would allow us, for example, to solve the following problems:
\begin{itemiz}
\item	If the control parameter is swept slowly in time, do we obtain a
	trustworthy representation of the bifurcation diagram?
\item	How do parameters, which cannot be controlled completely, but are
	subject to slow fluctuations, affect our modeling of the system?
\item	Consider a system subject to a slowly time--dependent driving
	force. Can we use the static bifurcation diagram (which is
	analytically more tractable) to gain some information on the
	time--dependent system? 
\end{itemiz}
To deal with this kind of questions, we should begin by understanding the
role of time scales in Physics.
 
%%%%%%%%%%%%%%%%%%%%%%%%%%%%%%%%%%%%%%%%%%%%%%%%%%%%%%%%%%%%%%%%%%%%%%%%%%%%%%%%%

\subsection{Slow--Fast Systems and Hysteresis}
\label{ssec_inhyst}

%%%%%%%%%%%%%%%%%%%%%%%%%%%%%%%%%%%%%%%%%%%%%%%%%%%%%%%%%%%%%%%%%%%%%%%%%%%%%%%%%

Physical systems are often characterized by one or several time scales. A
characteristic time might be the period of a typical periodic solution, or
the relaxation time to equilibrium. Let us consider a \DynSy\ with
characteristic time $T\sub1$, called the \defwd{fast system}\index{fast
system}, and couple it to another system with much larger characteristic
time $T\sub2 \gg T\sub1$, called the \defwd{slow system}\index{slow!system}.

Two particular situations are of interest:
\begin{enum}
\item	The evolution of the slow system is imposed from outside, and acts
	on the fast system as a slowly time--dependent parameter. For this
	purpose, it need not be governed by a differential equation. We call
	this coupled system an \defwd{adiabatic
	system}\index{adiabatic!system}. 
	
\item	The slow system is also a \DynSy, which is coupled to and influenced
	by the fast one. In this situation, we speak of a \defwd{slow--fast
	system}\index{slow--fast system}.
\end{enum}

As an illustration, let us imagine the following population model. In some
relatively small ecosystem, predators and prey reproduce, say, a couple of
times a year. Their populations have attained a cyclic regime, with a period
of a few years. Now the climate begins to change slowly, due for instance to
human impact, modifying the reproduction rate of the predator. This would be
an example of an adiabatic system, since the climate change is imposed from
outside. Another situation appears when, due to continual food consumption
by the prey, vegetation and {\em micro-climate} are slowly modified, changing
the reproduction rates in turn. This would be an example of a slow--fast
system. 

In this work, we are mainly interested in adiabatic systems. We believe,
however, that most results can be transposed to slow--fast systems (see
Chapter 4). 

\begin{figure}
 \centerline{
 \psfig{figure=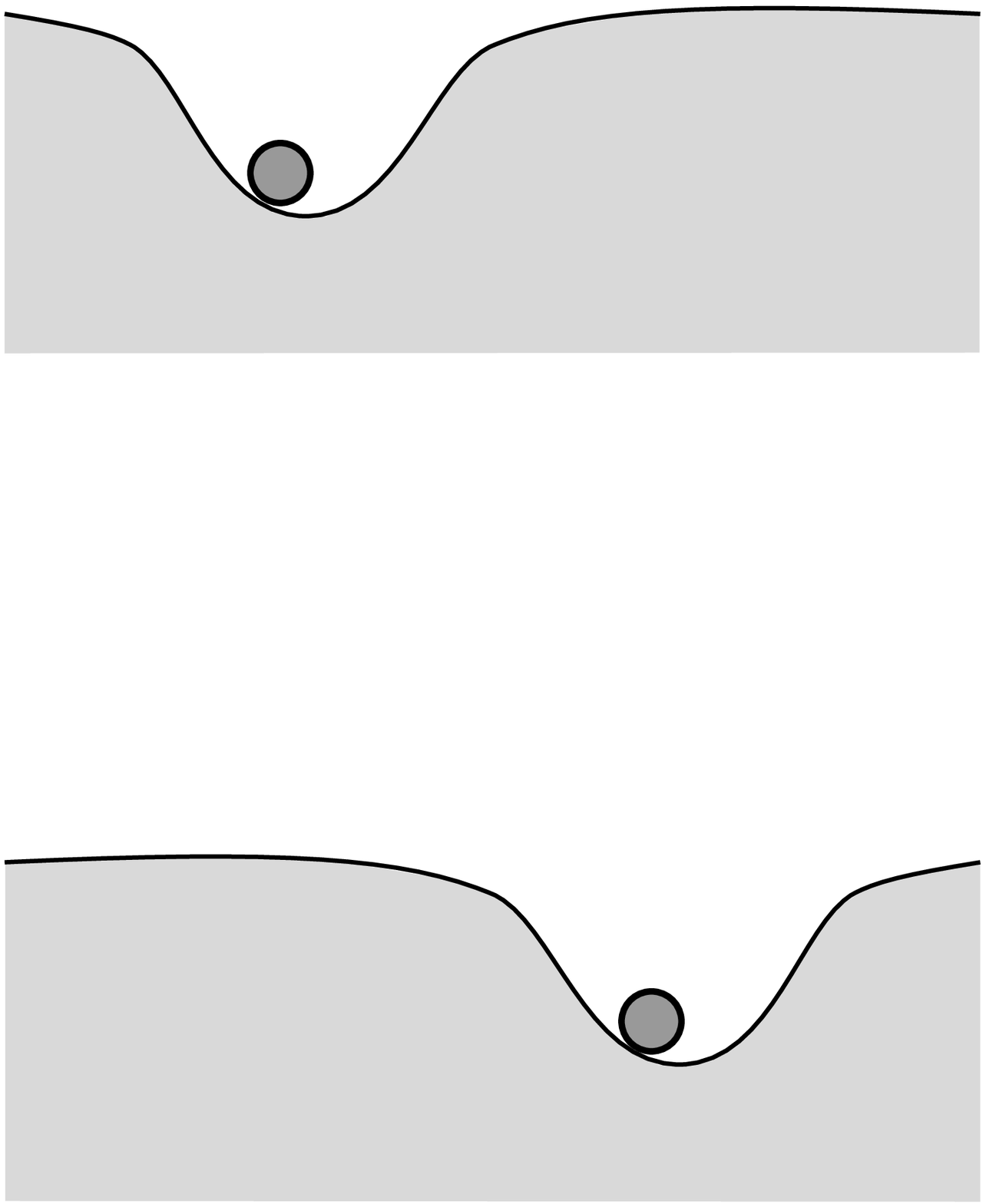,height=45mm,clip=t}
 \hspace{8mm}
 \psfig{figure=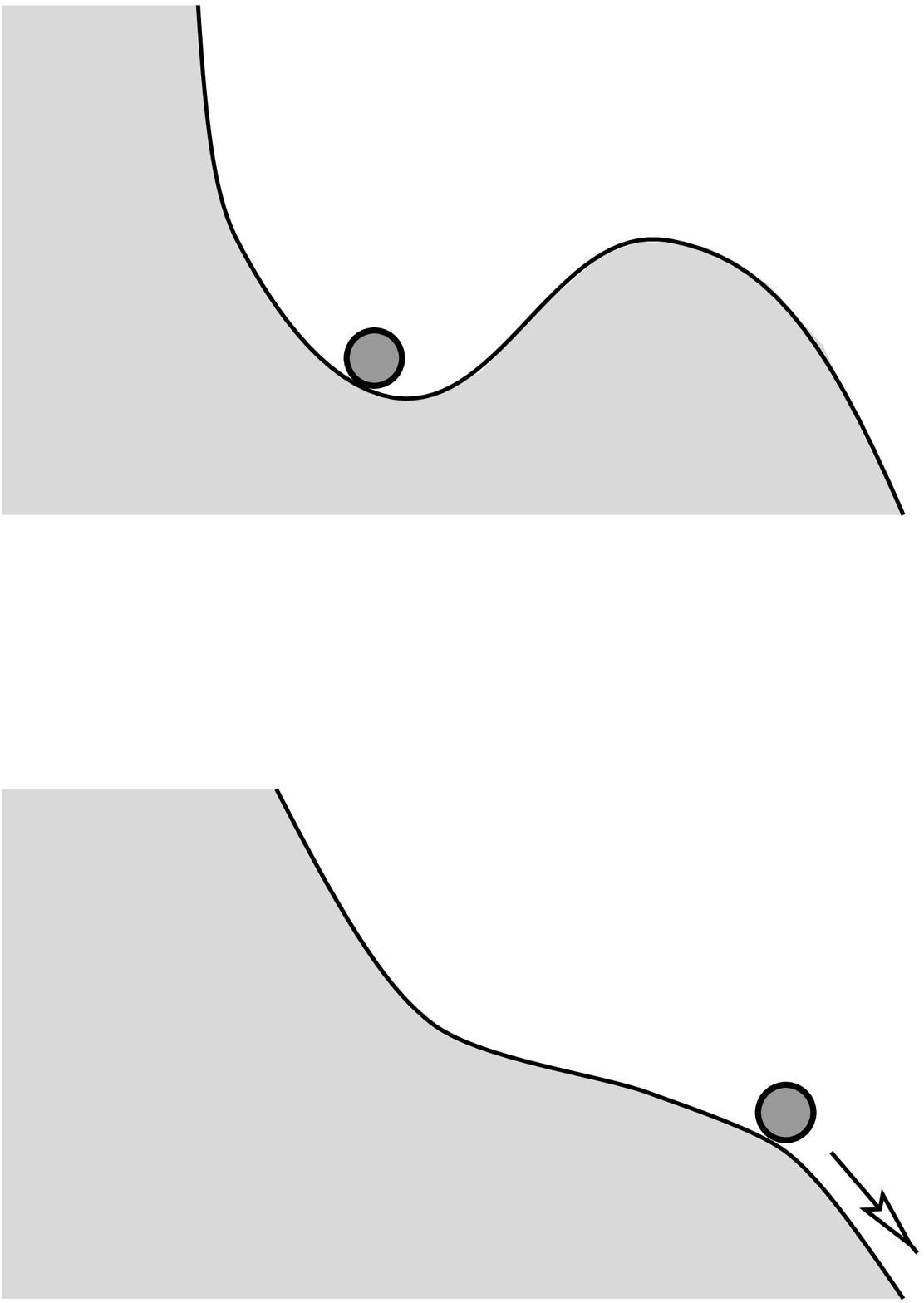,height=45mm,clip=t}
 \hspace{8mm}
 \psfig{figure=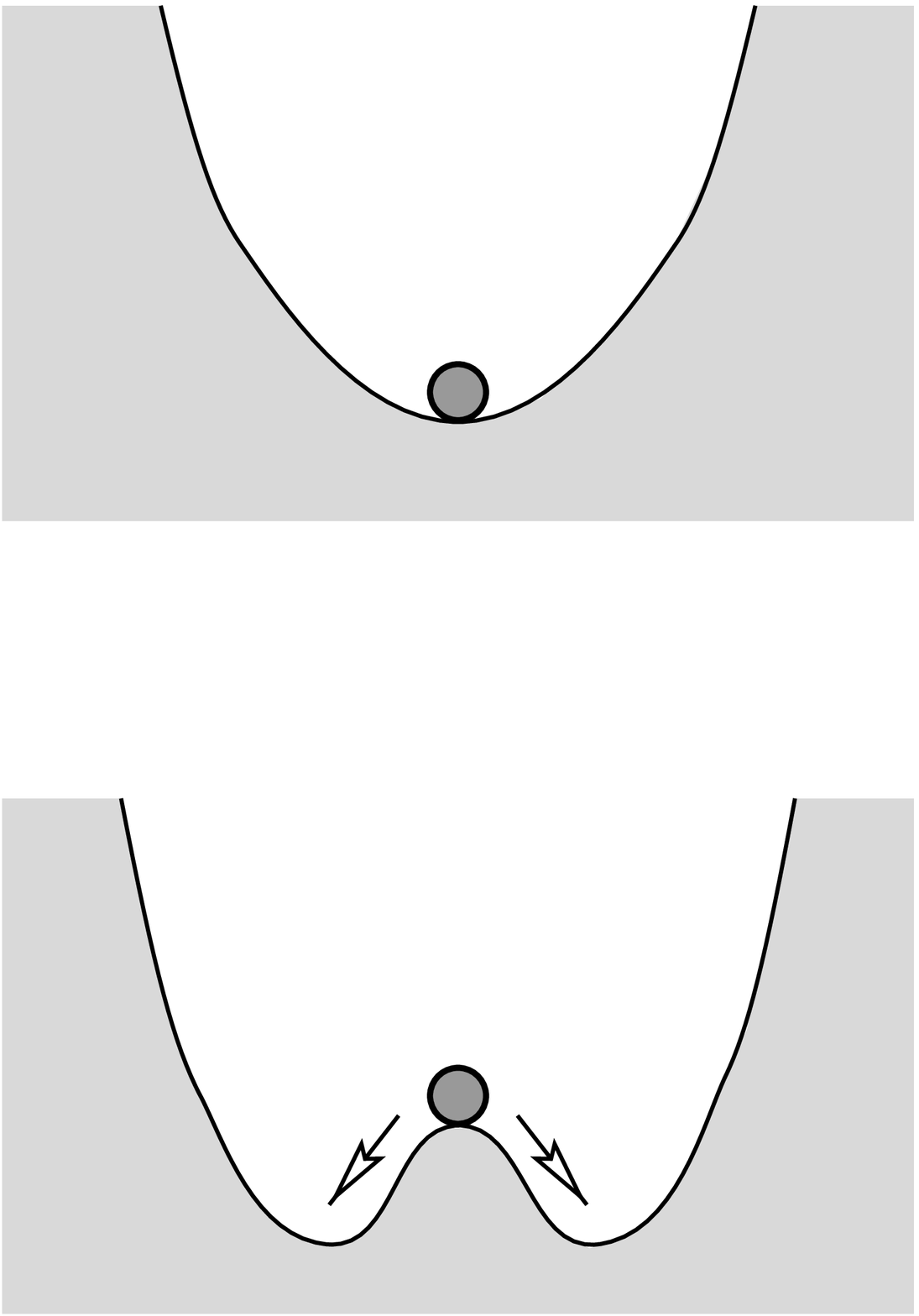,height=45mm,clip=t}
 \hspace{8mm}
 \psfig{figure=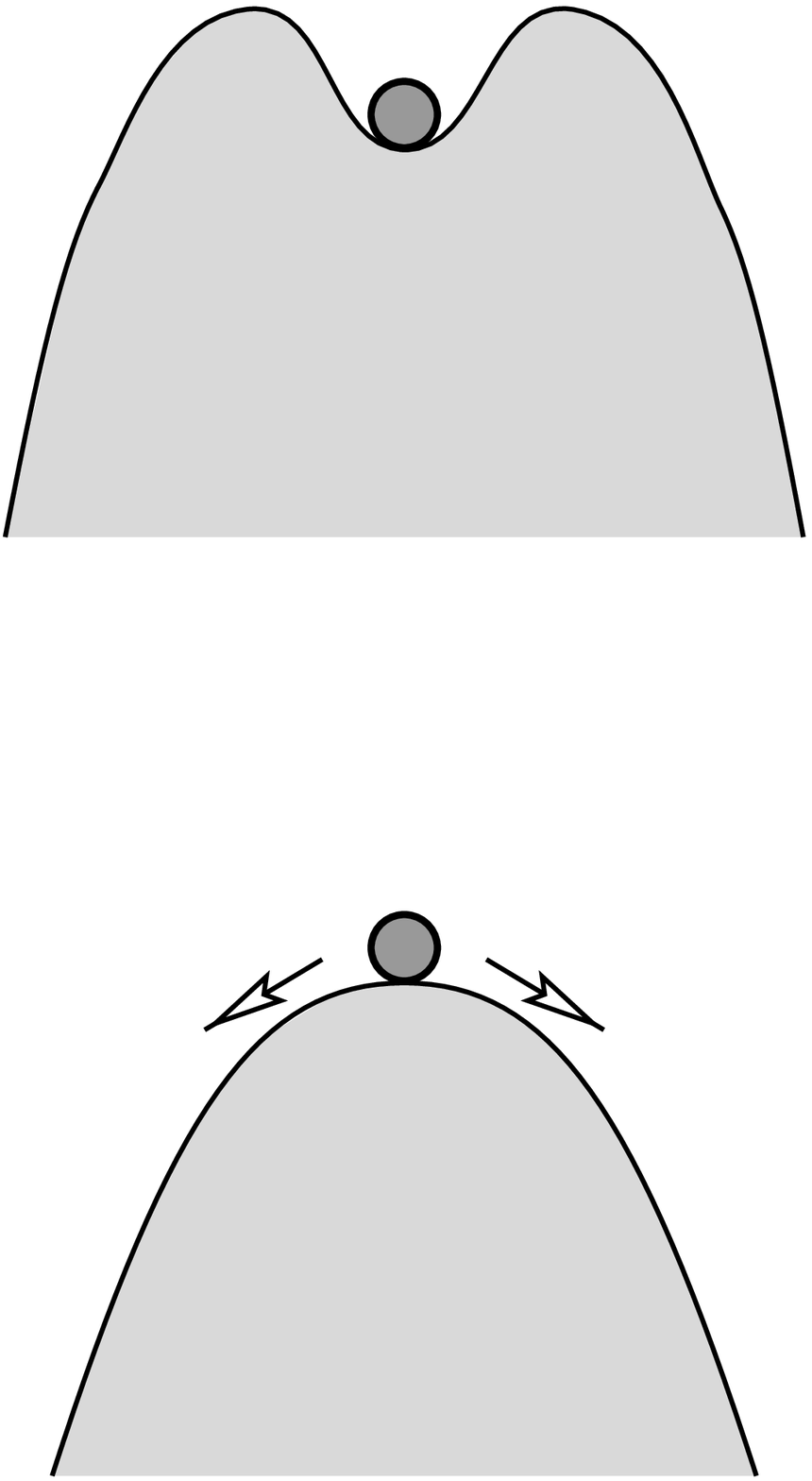,height=45mm,clip=t}
 }
 \figtext{
 	\writefig	0.0	2.7	\small{a}
 	\writefig	4.15	2.7	\small{b}
 	\writefig	8.4	2.7	\small{c}
 	\writefig	12.5	2.7	\small{d}
 }
% \vspace{2mm}
 \caption[Damped particle in a potential]
 {The damped motion of a particle in a slowly varying potential
 provides a simple example of adiabatic system. If the potential admits an
 isolated, slowly moving minimum, the particle will follow this well
 adiabatically (a). Bifurcations correspond to situations where this minimum
 interacts with other equilibrium points. For instance, the minimum may
 annihilate with a maximum (saddle--node bifurcation), and the particle
 leaves the vicinity of the bifurcation point (b). We may also have creation
 of two new equilibria (direct pitchfork bifurcation), so that the particle
 has to choose between its current, unstable position, and two potential
 wells (c). Or the minimum may disappear in favor of a maximum (indirect
 pitchfork bifurcation) (d).}
\label{fig_in2}
\end{figure}

What do we expect from the behaviour of an adiabatic system? To fix the
ideas, we can keep in mind the example of the motion of a damped particle,
in a slowly time--dependent potential. Let us first examine the case when
the static system (obtained by freezing the potential) admits a stable
stationary state (a potential minimum), depending smoothly on the
parameter. When the parameter is fixed, orbits starting in its \nbh\ will
relax to this equilibrium. When the parameter is swept slowly in time, it
is generally believed that the orbit will follow the equilibrium curve
adiabatically, \ie the particle will remain close (in a sense to be made
precise later) to the potential minimum. 

This behaviour has the following physical interpretation: in the adiabatic
limit, the asymptotic state will be identical with the static equilibrium
curve. In other words, the fast system is enslaved by the slow one, its
state being entirely determined by the value of the slow variables (\ie the
parameters).\footnote{To avoid a confusion due to terminology, we point out
that in thermodynamics, such a motion will be called \defwd{quasistatic}
rather than adiabatic.}

New phenomena arise when the equilibrium loses stability, a situation known
as \defwd{bifurcation}. Different scenarios are possible
(\figref{fig_in2}): the equilibrium may simply disappear, or it may become
unstable after interacting with one or several other equilibria. The
particle's motion depends a lot on the local structure of the bifurcation.
In some cases, it leaves the vicinity of the bifurcation point, until
reaching some other equilibrium or limit cycle. It may also follow a new
equilibrium branch created in the bifurcation, or even remain close to an
{\em unstable} equilibrium for some time, a phenomenon known as
\defwd{bifurcation delay}\index{bifurcation delay}, which can be
interpreted as metastability. These problems belong to the field of
\defwd{dynamic bifurcations}\index{dynamic!bifurcation}, which has received
much attention in recent years. 

\begin{figure}
 \centerline{\psfig{figure=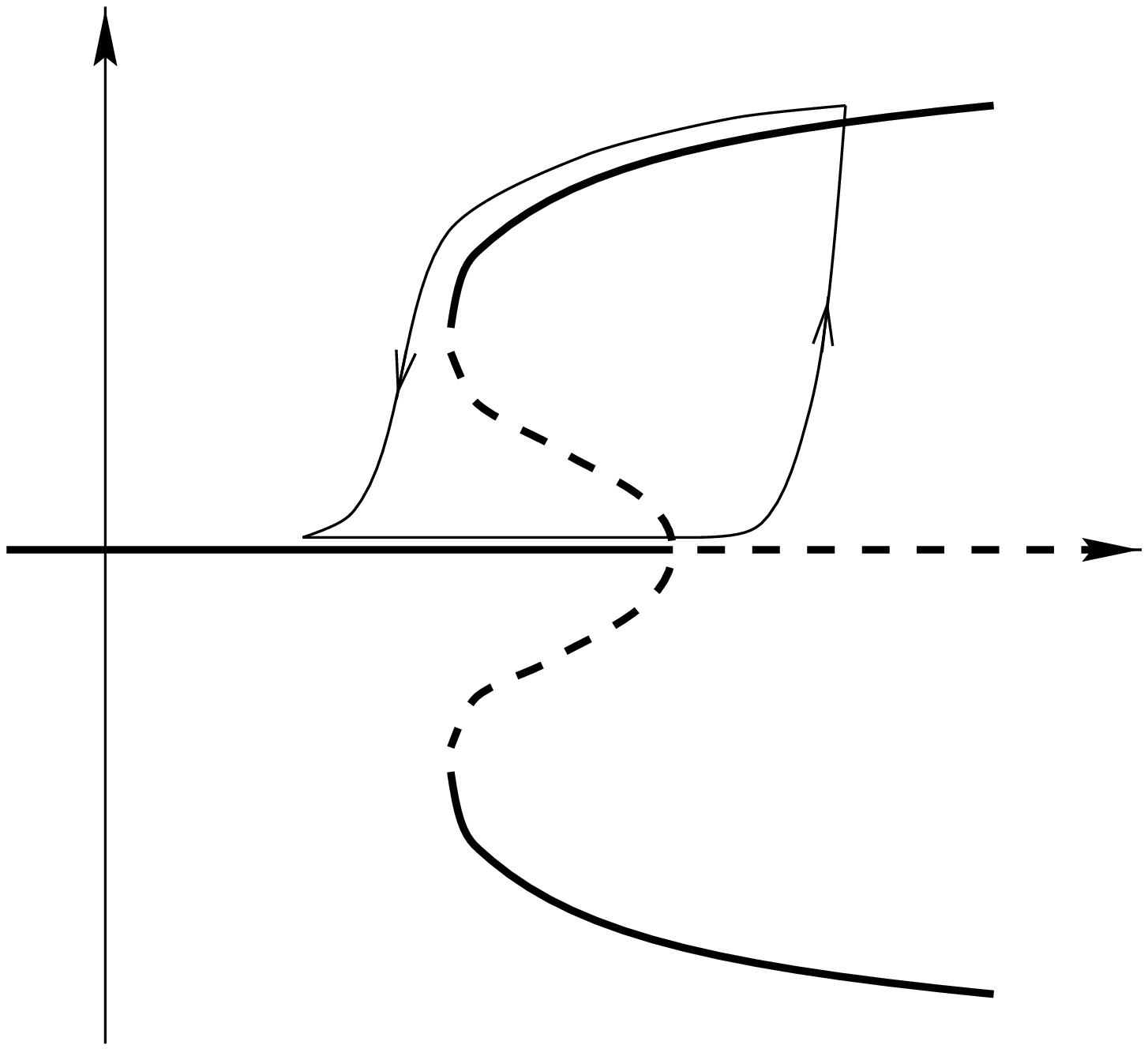,height=40mm,clip=t}
 \hspace{15mm}
 \psfig{figure=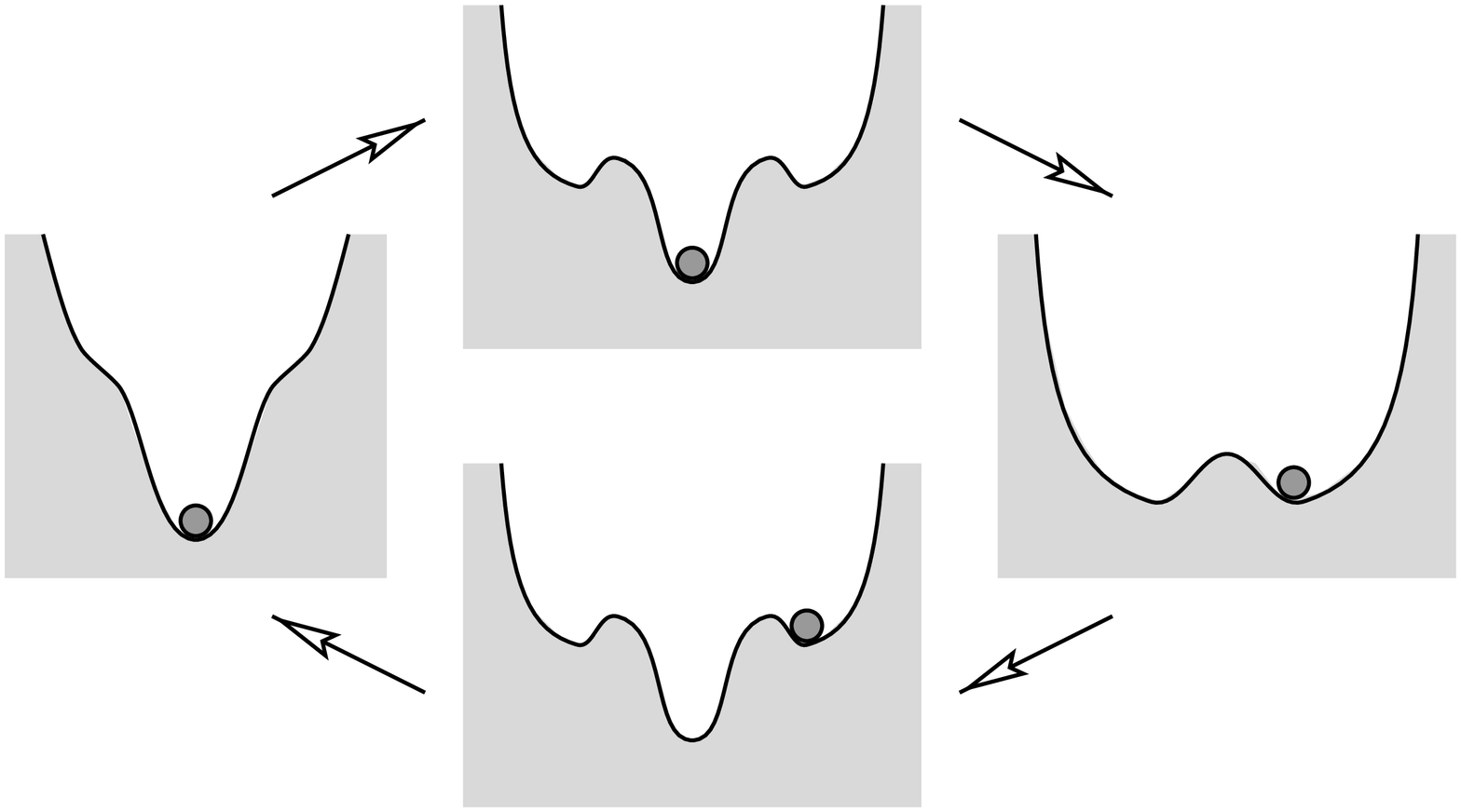,height=40mm,clip=t}}
 \figtext{
 	\writefig	3.7	2.0	\footnotesize{parameter}
 	\writefig	0.2	4.0	\footnotesize{state}
 }
 \caption[Example of hysteresis]
 {Example of a bifurcation diagram leading to hysteresis (a similar
 diagram  is found in \cite{Wiggins}). It can be seen as a combination of
 the bifurcations in \figref{fig_in2}b and d.  For increasing parameter,
 the solution follows the stable origin at least until the bifurcation (in
 fact, we will see that it may even follow the unstable origin for some
 time). When it finally reaches the new stable branch, and the parameter is
 decreased again, is stays close to this branch down to a smaller value of
 the parameter, hence describing a hysteresis cycle.}
\label{fig_in3}
\end{figure}

These local features of dynamics have a strong influence on global
properties. Let us focus on the situation when the parameter is varied
periodically in time. Without bifurcations, the solution will merely follow
the periodic motion of a stable equilibrium, independently of whether the
parameter is increasing or decreasing. The situation changes in presence of
bifurcations. It may happen, for instance, that the fast system follows a
different equilibrium branch for increasing or decreasing parameter. This
phenomenon is known as \defwd{hysteresis}\index{hysteresis}: the asymptotic
state depends not only on the present value of the parameter, but also on
its history (\figref{fig_in3}).

Hysteresis can be interpreted as the non--commutation of two limits, the
asymptotic and the adiabatic one. Mathematically, it is easier to take the
adiabatic limit first, which amounts to freezing the slow system. The motion
of the fast system is then governed by an autonomous (closed) equation, and
taking the asymptotic limit merely corresponds to analysing its equilibria
(or other attractors). 

This is, however, not the physically interesting information. We would like
instead to fix some small, but positive frequency of the parameter
variation, and study the asymptotic motion of the time--dependent fast
system (whatever this motion may be). {\em Then} we would like to determine
how this asymptotic motion behaves in the adiabatic limit, \ie when the
frequency of parameter variation goes to zero. 

Without bifurcation, the two limits can be taken in either order: the
asymptotic motion will approach a simple function of the parameter in the
adiabatic limit (see Example \ref{ex_in1} below). In presence of
bifurcations, hysteresis may occur. This is the central topic of this work.
To understand how hysteresis arises in adiabatic \DynSys, we first need to
develop methods which enable us to determine solutions for small, but
positive parameter sweeping rates. In particular, we have to understand if,
for a periodically varied parameter, these solutions tend asymptotically to
periodic ones, or if more complicated dynamics are possible. Then we will
be able to study their behaviour in the adiabatic limit.

%%%%%%%%%%%%%%%%%%%%%%%%%%%%%%%%%%%%%%%%%%%%%%%%%%%%%%%%%%%%%%%%%%%%%%%%%%%%%%%%%

\subsection{Historical Account}
\label{ssec_inhist}

%%%%%%%%%%%%%%%%%%%%%%%%%%%%%%%%%%%%%%%%%%%%%%%%%%%%%%%%%%%%%%%%%%%%%%%%%%%%%%%%%

We have no intention of giving here an exhaustive historical account of the
theory of \DynSys\ with multiple time scales. Besides the fact that such an
exposition would take many pages, we do not feel sufficiently well 
acquainted with the multiple aspects of this large domain to be able to
cite correctly the numerous researchers who contributed to one (or several)
of its facets. Instead, we would like to mention at this place the major
sources of inspiration of this work.

Research on adiabatic systems, hysteresis and related subjects appears to
have been pursued almost independently by mathematicians and physicists.
The former have been mostly interested in slow--fast systems, adiabatic
invariants and, more recently, in dynamic bifurcations and bifurcation
delay. The latter have rediscovered several times during this century the
importance of adiabatic systems. Recently, there has been renewed interest
in hysteresis appearing in lasers and magnets. Different models have been
considered, and studied  mainly by numerical methods.

\subsubsection*{Mathematics: slow--fast systems and bifurcation delay}
\index{action--angle variables}

Slow--fast systems have been studied almost since the beginning of
differential equations theory itself. They appear naturally in perturbed
integrable systems, where angle variables define the fast system, and
action variables the slow one. For instance, in the Solar System,
fast variables describe the motion of planets in their orbits, while slow
variables describe the spatial orientation of these orbits.\footnote{See
for instance Laskar's article in \cite{DD} for a non--technical
discussion.} When the interaction between planets is neglected, these
orbits are frozen in space, whereas they begin to deform slowly in time
when their interaction is taken into account.

Research on these systems has mainly focused on the dynamics of {\em slow}
variables. The method of averaging, for instance, aims at replacing the
dynamics of the slow variables by an effective equation, where the fast
variables have been averaged out \cite{ArG}. One often tries to construct
\defwd{adiabatic invariants}\index{adiabatic!invariant}, which are
functions on phase space remaining (almost) constant in time. A highlight of
this line of research is the celebrated Kolmogorov--Arnol'd--Moser (KAM)
theorem, which proves the existence of exact adiabatic invariants for some
initial conditions.

Adiabatic dynamics have been, for a long time, mainly studied in relation
with quantum mechanics \cite{Berry}. The \defwd{quantum adiabatic
theorem}\index{adiabatic!theorem} states that solutions of the slowly
time--dependent Schr\"odinger equation will adiabatically follow the
eigenspaces of the instantaneous Hamiltonian. Although this problem is
relatively old, rigorous proofs have been given only very recently
\cite{JKP}. Classical adiabatic systems (mostly linear ones) have been
studied in some detail by Wasow \cite{Wasow}.

An early result on nonlinear slow--fast systems is due to Pontryagin and 
Rodygin \cite{PR} in 1960. They showed that orbits of the fast system, which
start sufficiently close to a stable equilibrium or limit cycle, will follow
this attractor adiabatically. Problems involving bifurcations seem to have
been studied for the first time by Lebovitz and Schaar \cite{LS} in 1977.
They considered problems where two equilibrium branches exchange stability,
and showed that under some generic conditions, the orbit will follow a
stable branch after the bifurcation. 

In 1979, Haberman \cite{Ha} considered a class of one and
two--dimensional  problems. He introduced the notion of slowly varying
states, computed  as series in the adiabatic parameter, and studied in
particular jump phenomena (also  known as catastrophes) occurring near
saddle--node bifurcations.

The topic which would soon be given the name of \defwd{dynamic
bifurcations}\index{dynamic!bifurcation} developed rapidly in the second
half of the eighties. The importance of the \defwd{bifurcation
delay}\index{bifurcation delay} phenomenon in various physical situations
(lasers, neurons) was emphasized by Mandel, Erneux and co--workers
\cite{ME1,ME2,BER}, who derived an approximate formula for the delay time
using slowly varying states. This phenomenon, and the related problem of
\defwd{ducks}\index{duck} (also called \defwd{canards}) were then studied by
several mathematicians, using non--standard analysis (see \cite{Benoit} for
a summary of these works and a more detailed history).

A common feature of most of these works, including Wasow's, is that the
authors try to construct particular solutions as series in the adiabatic
parameter. The problem is, however, that these series are in general {\em
not} convergent. A naive treatment of such equations may therefore yield, in
some cases, incorrect results. In order to obtain the right answers with
these methods (for instance the fact that there exists a maximal value for
the bifurcation delay), one has to use rather elaborate techniques, as
resummation of divergent series (see \cite{Benoit}, in particular the
articles by Diener and Diener, and by Canalis--Durand).

An entirely new direction to treat these problems was initiated by Neishtadt
\cite{Ne1,Ne2}. Returning to the old technique of successive changes of
variables, but combined with estimations inspired by Nekhoroshev, he was
able to prove rigorously the existence of a bifurcation delay. Moreover,
with the help of a technique involving deformation of an integration path
into the complex plane, he could give an explicit lower bound to the
delay time. Diener and Diener \cite{Benoit} have examined under which
generic conditions this formula gives an upper bound as well.

Recently, these results have been generalized to the case of a periodic
orbit undergoing Hopf bifurcation [NST].

\begin{table}
\index{hysteresis!cycle!scaling}\index{scaling!of hysteresis cycle}
\vspace{10mm}
\begin{center}
\begin{tabular}{|l|l|l|}
\hline
	 & {\bf System}		& {\bf Scaling of $\cA$} 		\\
\hline	
\hline	
Experiments
			& Iron \cite{Steinmetz}	
			& $H\sub0^{1.6}$				\\
			& Fe/Au film \cite{HeWa}	
			& $H\sub0^{0.59}\Omega^{0.31}$			\\
			& Co/Cu film \cite{JYW}	
			& $\cA\sub0 + 
			  (H\sub0^2-\crit{H}^2)^{0.34}\Omega^{0.66}$	\\
			& Fe/W film \cite{SE}	
			& $H\sub0^{0.25}\Omega^{0.03}$			\\
\hline	
Numerical simulations
			& $(\Phi^2)^2$ model \cite{RKP}		
			& $H\sub0^{0.66}\Omega^{0.33}$	 		\\
			& $(\Phi^2)^2$ model \cite{ZZS}		
			& $\Omega^{0.5}$	 			\\
			& $(\Phi^2)^3$ model \cite{ZZS}		
			& $\cA\sub0 + \Omega^{0.7}$			\\
			& Ising 2D Monte--Carlo \cite{LP}		
			& $H\sub0^{0.46}\Omega^{0.36}$	 		\\
			& Ising 2D Monte--Carlo \cite{ZZL}		
			& $\cA\sub0 + \Omega^{0.36}$			\\
			& Ising 2D Monte--Carlo \cite{AC1}		
			& $H\sub0^{0.7}\Omega^{0.36}$			\\
			& Ising 3D Monte--Carlo \cite{AC1}		
			& $H\sub0^{0.67}\Omega^{0.45}$			\\
			& Cell--dynamical system \cite{ZZL}		
			& $\cA\sub0 + \Omega^{0.66}$			\\
			& Mean field \cite{LZ}		
			& $\cA\sub0 + H\sub0^{0.66}\Omega^{0.66}$	\\
\hline	
Analytical arguments
			& Mean field \cite{Jung}		 	
			& $\cA\sub0 + 
			  (H\sub0^2-\crit{H}^2)^{1/3} \Omega^{2/3}$	\\
			& $(\Phi^2)^2$ model \cite{DT}		 	
			& $H\sub0^{1/2} \Omega^{1/2}$			\\
			& $(\Phi^2)^2$ model \cite{SD}		 	
			& $H\sub0^{1/2} \Omega^{1/2}$			\\
			& $(\Phi^2)^2$ model \cite{ZZ2}		 	
			& $\Omega^{1/2}$				\\
			& Ising dD \cite{SRN}		 	
			& $\abs{\ln\Omega}^{-1/(d-1)}$			\\
\hline
\end{tabular}
\end{center}
\vspace{2mm}
\caption[Scaling of hysteresis cycles area]
{Some results on the scaling behaviour of the area $\cA$ enclosed by a
hysteresis cycle, as a function of magnetic field amplitude $H\sub0$ and
frequency $\Omega$. Recent experiments were made with ultrathin films. 
Numerical Monte--Carlo simulations have been carried out on the
two--dimensional (2D) and three-dimensional (3D) Ising model with Glauber
dynamics. Other numerical experiments concern the Langevin equation in a
Ginzburg--Landau $(\Phi^2)^2$ or $(\Phi^2)^3$ potential with $\O{N}$
symmetry. In the large $N$ limit, the noise can be eliminated from the
equation, and one obtains deterministic ODE. The proposed exponents differ
a lot from one experiment to another. In particular, it is not clear
whether the area should go to zero, or to a finite limit $\cA\sub0$ when
$\Omega\to 0$. It is amusing to note that results of one experiment
\cite{JYW} could  be fitted on the mean field result \cite{Jung}, while
another one \cite{HeWa} was fitted on results of the $(\Phi^2)^2$ model
studied in \cite{RKP}. Although the mean--field studies in \cite{Jung} and
\cite{LZ} predict the same $\Omega$--dependence, they do not agree on the
$H\sub0$--dependence. In fact, we will show that both laws are incorrect.}
\label{tab_in2}
\vspace{10mm}
\end{table}

\subsubsection*{Physics: hysteresis and scaling laws}

Research on hysteresis has been pursued by physicists, almost independently
of mathematicians, and mostly with numerical methods. For a long time,
the standard model for hysteretic phenomena has been the Preisach model
\cite{May,MNZ}. This model, however, is artificial and provides no
derivation of hysteresis from microscopic principles. 

Interest in microscopic models of magnetic hysteresis was renewed in 1990,
by an important article by Rao, Krishnamurthy and Pandit \cite{RKP}. They
analyse numerically two models, an Ising model with Monte--Carlo dynamics,
and a continuous model with $\O{N}$ symmetry in the large $N$ limit. They
proposed in particular that the area $\cA$ enclosed by the hysteresis cycle
should scale with the amplitude $H\sub0$ of the magnetic field and its
frequency $\Omega$ according to the power law $\cA \ord
H\sub0^{\alpha}\Omega^{\beta}$, where $\alpha \ord 0.66$ and $\beta \ord
0.33$ for small frequency and amplitude.

This work inspired a large number of articles trying to exhibit scaling
laws for hysteresis cycles. In the case of a laser system \cite{Jung},
analytical arguments showed that a one--dimensional model equation admits a
hysteresis cycle with area $\cA(\Omega) \ord \cA(0) + \Omega^{2/3}$. The
discrepancy between this result and the one in \cite{RKP} lead in following
years to some controversy \cite{Ra}.

Still in the year 1990, a numerical study of a mean field approximation of
the Ising model introduced the concept of a \defwd{dynamic phase
transition}\index{dynamic!phase transition}\index{phase!transition!dynamic}
\cite{TO}: regions with zero and non--zero average magnetization by cycle
are separated by a transition line in the
temperature--magnetic-field-amplitude plane. 

These papers were followed by various numerical simulations (on lattice
models and continuous ones) and experiments, which proposed new sets of
exponents. We show some of them in \tabref{tab_in2}. The trouble is that
even for one and the same model, these exponents differ widely from one
experiment to the other.

There have been several attempts to derive these exponents analytically.
Relatively simple systems, like lasers, seem to be described satisfactorily
by one--dimensional equations, as shown in \cite{Hohl,GBS} which extend
results in \cite{Jung}. However, for magnetic systems, no satisfactory
explanation has been obtained. Some analytical arguments, using rescaling
\cite{SD} or renormalization \cite{ZZ2} seem to indicate that the area
should scale as $\cA \ord H\sub0^{1/2}\Omega^{1/2}$. Various explanations
have been proposed for these discrepancies, for instance logarithmic
corrections \cite{DT}. 

In fact, it is not clear at all whether the area should really follow a
power law \cite{SRN}. It depends probably in a crucial way on the detailed
dynamics of droplets during magnetization reversal. At any rate,
understanding how these scaling laws may appear in the model equations would
be a good criterion to test their adequacy against real physical systems.
Recently, several authors have introduced other models, including quantum
effects \cite{BDS}; they have also become interested in other indicators,
like pulse susceptibility \cite{AC3}.

%%%%%%%%%%%%%%%%%%%%%%%%%%%%%%%%%%%%%%%%%%%%%%%%%%%%%%%%%%%%%%%%%%%%%%%%%%%%%%%%%

\section{Mathematical Formulation}
\label{sec_inmath}

%%%%%%%%%%%%%%%%%%%%%%%%%%%%%%%%%%%%%%%%%%%%%%%%%%%%%%%%%%%%%%%%%%%%%%%%%%%%%%%%%

%%%%%%%%%%%%%%%%%%%%%%%%%%%%%%%%%%%%%%%%%%%%%%%%%%%%%%%%%%%%%%%%%%%%%%%%%%%%%%%%%

\subsection{Adiabatic Systems and Slow--Fast Systems}
\label{ssec_inadiab}

%%%%%%%%%%%%%%%%%%%%%%%%%%%%%%%%%%%%%%%%%%%%%%%%%%%%%%%%%%%%%%%%%%%%%%%%%%%%%%%%%

We will consider dynamical systems described by ordinary differential
equations of the form
\begin{equation}
\label{in1}
\dtot{x}{t} = f(x,\lambda),
\end{equation}
where $x\in\R^n$ is the vector of dynamic variables, and $\lambda\in\R^p$ is
a set of parameters. We shall assume that $f$ is a function of class
$\diff{2}$ at least. 

The slow variation of parameters is described by a function $G(\eps t)$,
where $0<\eps\ll 1$ is the \defwd{adiabatic
parameter}\index{adiabatic!parameter}:
\begin{equation}
\label{in2}
\dtot{x}{t} = f(x, G(\eps t)).
\end{equation}
This formulation should be interpreted as follows: $f(x,\lambda)$ and
$G(\tau)$ are given functions, fixed once and for all,\footnote{One may, in
fact, allow for an $\eps$--dependence of $f$, provided $f$ behaves smoothly
(in some sense)  in the limit $\eps\to 0$, see Section 4.1.}
and we would like to understand the behaviour of \eqref{in2} in the
\defwd{adiabatic limit}\index{adiabatic!limit} $\eps\to 0$. For instance,
$G(\eps t) = \sin(\eps t)$ would describe a periodic variation of the
parameter, with small frequency $\eps$.

The adiabatic limit should be taken with some care. If we naively replace
$\eps$ by 0 in \eqref{in2}, we obtain the autonomous system $\dtot{x}{t} =
f(x,G(0))$. This is due to the fact that with respect to the slow time
scale, we have zoomed on a particular instant. This is not what we are
interested in: it is more natural, for our purpose, to study the system on
the slow time scale of parameter variation. We do that by introducing a
\defwd{slow time}\index{slow!time} $\tau = \eps t$, so that \eqref{in2} can
be rewritten
\begin{equation}
\label{in3}
\eps\dtot{x}{\tau} \bydef \eps\dot{x} = f(x,G(\tau)).
\end{equation}
We call this equation an \defwd{adiabatic system}\index{adiabatic!system}.
In the adiabatic limit, it reduces to the algebraic equation $f(x,G(\tau)) =
0$. We will see that although this limit is singular, it is less
problematic to analyse than for \eqref{in2}. 

By contrast, a \defwd{slow--fast system}\index{slow--fast system} is
described by a set of coupled ODE of the form
\begin{equation}
\label{in4}
\begin{split}
\eps\dot{x} &= f(x,y) \\
\dot{y} &= g(x,y).
\end{split}
\end{equation}
In some circumstances, adiabatic and slow--fast systems are equivalent and
may be transformed into one another. For instance, if $G(\tau)$ is the
solution of a differential equation $\dot{y} = g(y)$, the adiabatic system
\eqref{in3} can be transformed into a slow--fast system. If $\lambda\in\R$,
this transformation is only possible for \monoton\ $G(\tau)$. There are
other ways to write \eqref{in3} as a vector field, for instance by
considering the slow time $\tau$ as a dynamic variable (see next
subsection). In some particular cases, it may be helpful to introduce
additional variables, for instance $G(\tau) = \sin\tau$ is a solution of
$\dot{y}=z$, $\dot{z}=-y$. 

On the other hand, if $g(x,y)$ depends only on $y$, the slow--fast system
\eqref{in4} is equivalent to the adiabatic system \eqref{in3}, with
$G(\tau)$ given by the solution of $\dot{y} = g(y)$. If $g$ depends on $x$
as well, this reduction is not possible, but one can sometimes construct a
solution in the following way: in first approximation, $x$ is related to
$y$ by the algebraic equation $f(x,y)=0$. If this equation admits a unique
solution $x=\fix{x}(y)$, $y(\tau)$ may be approximated by a solution of the
equation $\dot{y} = g(\fix{x}(y),y)$, which can be used in turn to estimate
corrections to the solution $x(\tau) \simeq \fix{x}(y(\tau))$.

%%%%%%%%%%%%%%%%%%%%%%%%%%%%%%%%%%%%%%%%%%%%%%%%%%%%%%%%%%%%%%%%%%%%%%%%%%%%%%%%%

\subsection{Adiabatic Systems and Vector Fields}
\label{ssec_invect}

%%%%%%%%%%%%%%%%%%%%%%%%%%%%%%%%%%%%%%%%%%%%%%%%%%%%%%%%%%%%%%%%%%%%%%%%%%%%%%%%%

We can exploit the similarities with slow--fast systems to obtain valuable
informations on the solutions of the adiabatic system \eqref{in3} without
any analytical calculation. This is done by using geometric properties of
vector fields. For simplicity, we consider the case of a scalar parameter
$\lambda\in\R$.

It is always possible to write \eqref{in3} as a vector field by considering
the slow time $\tau$ as a dynamic variable:
\begin{equation}
\label{in5}
\begin{split}
\dtot{x}{t} &= f(x,G(\tau))\\
\dtot{\tau}{t} &= \eps.
\end{split}
\end{equation}
A major drawback is that this vector field has no singular points. One can
however deduce some general properties of the flow. When $f(x,G(\tau))\neq
0$, orbits have a large slope of order $\eps^{-1}$, due to the short
characteristic time of the fast variable $x$. On the other hand, when
$f(x,G(\tau))=0$, the vector field is parallel to the $\tau$--axis. In fact,
in a \nbh\ of order $\eps$ of an equilibrium branch, the motion of the fast
variable becomes slow: it is the region where \defwd{adiabatic
solutions}\index{adiabatic!solution} (also known as slowly varying states)
may exist.

In the case $x\in\R$ ($n=1$), the form of the vector field \eqref{in5}
imposes strong constraints on the solutions. It is possible to show, using
only geometric arguments, that some solutions will remain in the \nbh\ of
equilibrium branches of $f$ (\figref{fig_in4}). We will see that this
property can be generalized to the $n$--dimensional case.

There are two particular classes of functions $G(\tau)$ for which it is
possible to say more:

\begin{figure}[tb]
 \centerline{\psfig{figure=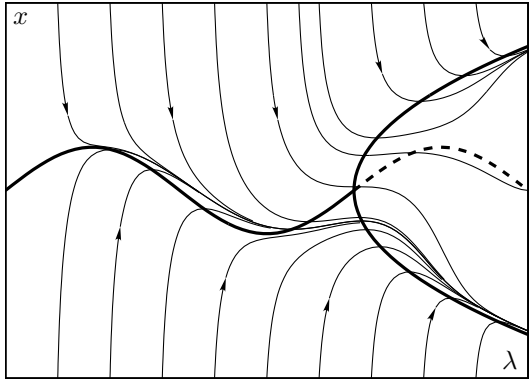,height=50mm,clip=t}}
 \figtext{
 	\writefig	4.0	5.2	$x$
 	\writefig	10.5	0.6	$\lambda$
 }
 \caption[Solutions of $\eps\dot{x} = f(x,\tau)$]
 {Solutions of the equation $\eps\dot{x} = f(x,\lambda)$, for
 $\lambda(\tau)=\tau$ [here, the function $f$ is given by
 $f(x,\lambda)=(x-\sin(\pi\lambda)/2)(\lambda-x^2)$]. The curves on which
 $f(x,\lambda)=0$ (thick lines) delimit regions where the vector field has
 positive or negative slope. This imposes geometrical constraints on the
 solutions. In the left half of the picture, there exists a stable
 equilibrium branch. Solutions lying above this branch are decreasing,
 while those lying below are increasing. From this construction, one can
 already deduce existence of adiabatic solutions, remaining close to the
 equilibrium. For a special parameter value, there is a bifurcation: the
 equilibrium becomes unstable, and new stable branches are created. In this
 case, adiabatic solutions coming from the left follow the lower branch.}
\label{fig_in4}
\end{figure}

\subsubsection*{Monotonous case}

If $G(\tau)$ is strictly \monoton, it admits an inverse function $G^{-1}$.
We may thus use $\lambda = G(\tau)$ as a dynamic variable, giving
\begin{equation}
\label{in6}
\begin{split}
\eps\dot{x} &= f(x,\lambda) \\
\dot{\lambda} &= g(\lambda) \defby G'(G^{-1}(\lambda)).
\end{split}
\end{equation}
If $G'(\tau)$ goes to zero in some limit, fixed points may appear in the
vector field. 

\begin{figure}
 \centerline{
 \psfig{figure=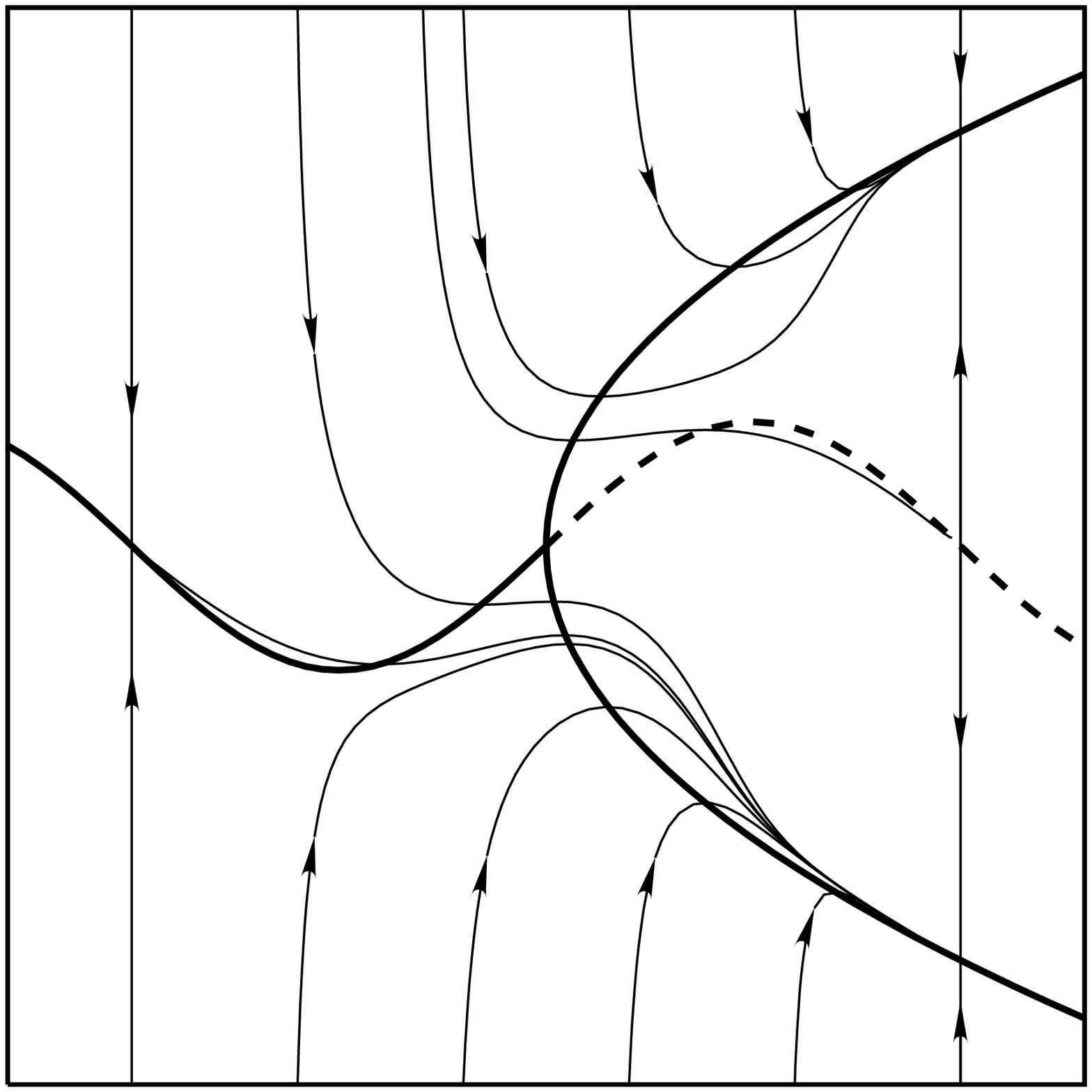,height=50mm,clip=t}
 \hspace{18mm}
 \psfig{figure=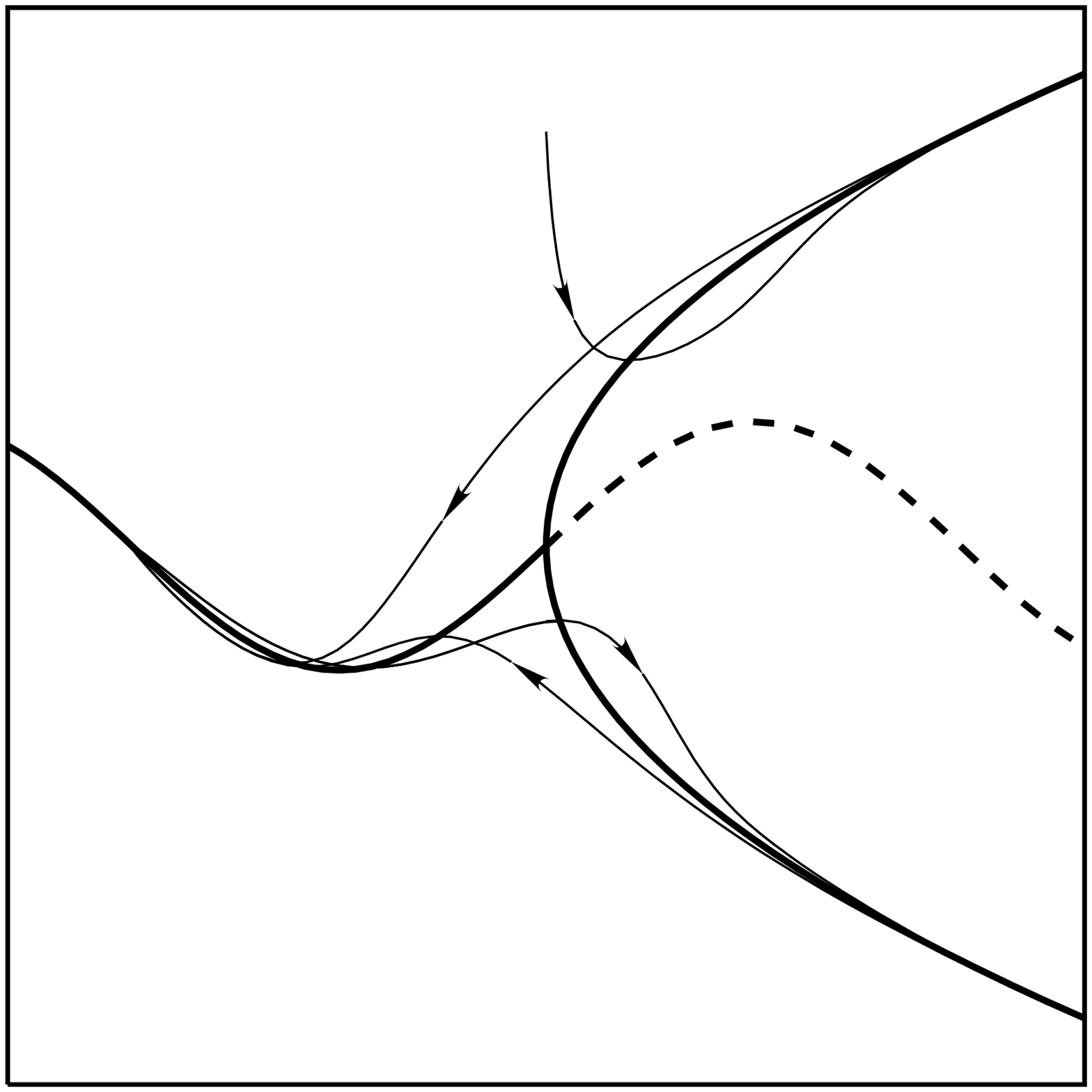,height=50mm,clip=t}
 }
 \figtext{
 	\writefig	0.9	5.2	a
	\writefig	7.9	5.2	b
 	\writefig	1.5	5.2	$x$
	\writefig	8.5	5.2	$x$
 	\writefig	6.5	0.55	$\lambda$
	\writefig	13.5	0.55	$\lambda$
 }
 \caption[Solutions of $\eps\dot{x} = f(x,\thyp\tau)$ and $\eps\dot{x} =
 f(x,\sin\tau)$]
 {Same equation as in \figref{fig_in4}, but with (a) $\lambda(\tau) =
 \thyp(\tau)$ and (b) $\lambda(\tau) = \sin(\tau)$. In (a), the system
 admits hyperbolic fixed points at $(\pm 1,0)$, and stable nodes at $(1,\pm
 1)$, which define the asymptotic states. The stable manifold of $(1,0)$
 delimits the basins of attraction. In this case, all trajectories reach
 the lower equilibrium. In (b), during the first cycle the solution follows
 the upper branch, which is still a transient motion.  From the next cycle
 on, it is attracted by a periodic orbit following the lower branches.}
\label{fig_in5}
\end{figure}

Consider for instance the case $G(\tau) = \thyp\tau$. Then $g(\lambda) =
1-\lambda^2$ vanishes at $\lambda = \pm 1$. As $\tau$ goes to infinity
($\lambda\to 1$), trajectories will be attracted by stable fixed points of
$f(x,1)$. We conclude that if $\lambda$ moves infinitely smoothly from an
initial to a final value, we can construct a smooth transformation which
compactifies phase space, and in this way the asymptotic limit
$\tau\to\infty$ can be properly defined (\figref{fig_in5}a).

%\newpage

\subsubsection*{Periodic case}

Assume $G(\tau)$ is periodic, say $G(\tau) = \sin\tau$. We can write the
adiabatic system in the form \eqref{in5}, with the particularity that
$\tau$ can be considered as a periodic variable (\ie the phase space has
the topology of a cylinder). Since the flow is transverse to every plane
$\tau = \text{constant}$, dynamics can be characterized by the
\defwd{Poincar\'e section}\index{Poincar\'e!section} at $\tau=0$ (say), and
its \defwd{Poincar\'e map}\index{Poincar\'e!map} $T: x(0) \mapsto x(1)$. In
particular, periodic orbits correspond to fixed points of $T$. In the
one--dimensional case, this fact can be used to prove that every orbit is
either periodic, or attracted by a periodic orbit.

Of course, to study hysteresis properties, we would like to go back to
$(\lambda,x)$--variables, which is done by ``wrapping'' the
$(\tau,x)$--space (\figref{fig_in5}b). Some information can also be gained
by using a representation of the form \eqref{in5} on each interval in
which  $G(\tau)$ is \monoton. One should, however, pay attention to the
fact that this transformation introduces artificial singularities in the
vector field at those points where $G'(\tau)$ vanishes.

%%%%%%%%%%%%%%%%%%%%%%%%%%%%%%%%%%%%%%%%%%%%%%%%%%%%%%%%%%%%%%%%%%%%%%%%%%%%%%%%%

\subsection{Some Simple Examples}
\label{ssec_inex}

%%%%%%%%%%%%%%%%%%%%%%%%%%%%%%%%%%%%%%%%%%%%%%%%%%%%%%%%%%%%%%%%%%%%%%%%%%%%%%%%%

Let us return to the example of the damped motion of a particle in a
potential, which is described by an equation of the form
\begin{equation}
\label{in7}
\dtot{^2x}{t^2} + \gamma\dtot{x}{t} + \dpar{\Phi}{x}(x,\lambda) = 0.
\end{equation}
We will show in Chapter 3 that for sufficiently large friction,
this system is governed by the one--dimensional equation
\begin{equation}
\label{in8}
\dtot{x}{t} = f(x,\lambda) \simeq -\dpar{\Phi}{x}(x,\lambda).
\end{equation}
Thus, any one--dimensional system of the form 
\begin{equation}
\label{in9}
\eps\dot{x} = f(x,\lambda(\tau))
\end{equation}
may be interpreted as describing the overdamped motion of a particle in a
slowly varying potential $\Phi(x,\lambda(\tau))$, with $f \simeq
-\sdpar{\Phi}{x}$. Let us examine some particular cases, in order to
illustrate the previously discussed concepts.

\begin{figure}
 \centerline{
 \psfig{figure=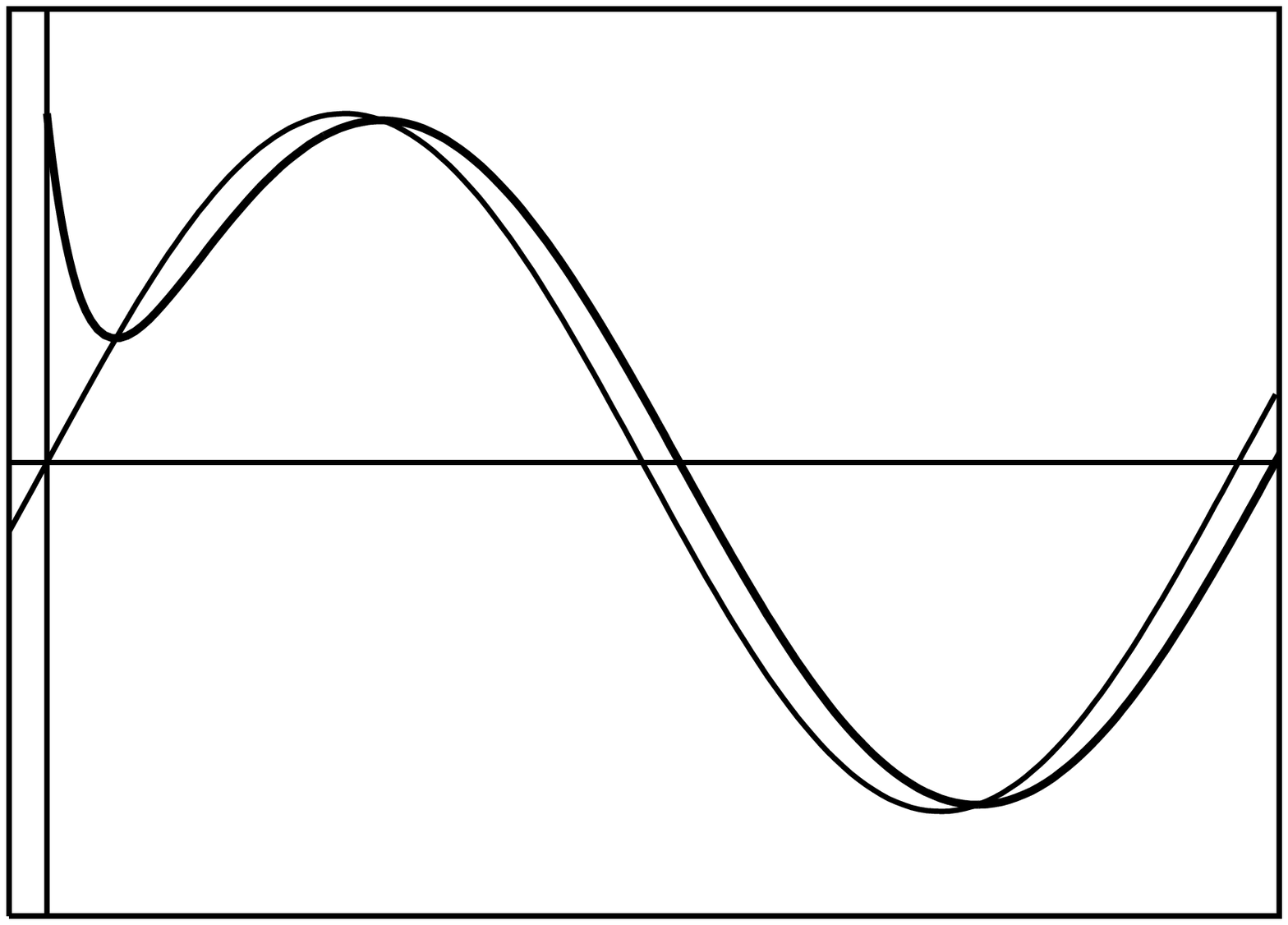,height=50mm,clip=t}
 \hspace{10mm}
 \psfig{figure=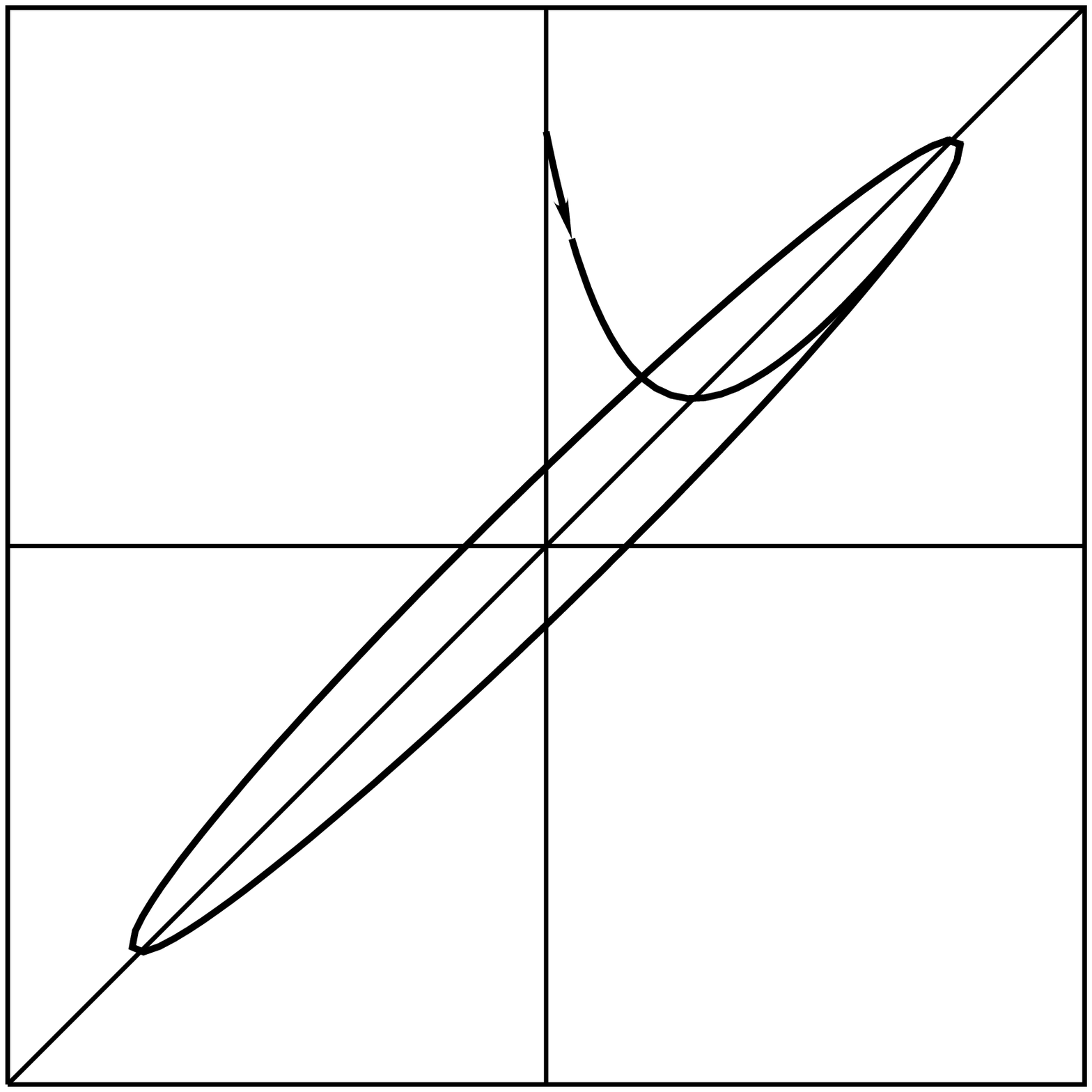,height=50mm,clip=t}
 }
 \figtext{
 	\writefig	0.5	5.2	a
	\writefig	8.6	5.2	b
 	\writefig	1.2	5.2	$x$
	\writefig	11.6	5.2	$x$
 	\writefig	7.3	3.1	$\tau$
	\writefig	13.6	3.1	$\lambda$
 }
 \caption[Solutions of the equation $\eps\dot{x} = -x + \sin\tau$]
 {Solutions of the equation $\eps\dot{x} = -x + \sin\tau$ of Example
 \ref{ex_in1}. (a) After a short transient, the solution $x(\tau)$ (thick
 line) follows adiabatically the forcing $\sin\tau$ (thin line), with a
 phase shift of order $\eps$. (b) The ``Lissajous plot'' of this solution
 in the $(\lambda,x)$--plane is attracted by an ellipse, at a distance of
 order $\eps$ from the line $x=\lambda$. This cycle encloses an area of
 order $\eps$ which vanishes in the adiabatic limit, thus we do not
 consider it as a hysteresis cycle.}
\label{fig_in6}
\end{figure}

\begin{example}
\label{ex_in1}
Consider the equation 
\begin{equation}
\label{in10}
\eps\dot{x} = -x + \lambda(\tau), 
\qquad	\lambda(\tau) = \sin\tau,
\end{equation}
which describes the motion of an overdamped, adiabatically forced harmonic
oscillator. It can be solved explicitly, with the result
\begin{equation}
\label{in11}
x(\tau) = \biggbrak{x(0) + \frac{\eps}{1+\eps^2}} \e^{-\tau/\eps} 
+ \frac{1}{1+\eps^2} (\sin\tau - \eps\cos\tau).
\end{equation}
The second term is a periodic particular solution of \eqref{in10}. It
follows the forcing $\lambda(\tau)$ with a phase shift of order $\eps$
(\figref{fig_in6}a). This is precisely what we call an \defwd{adiabatic
solution}\index{adiabatic!solution}, since it remains in a \nbh\ of order
$\eps$ of the static equilibrium $x=\lambda(\tau)$. In the
$(\lambda,x)$--plane, it is represented by an ellipse with width of order
$\eps$ (which can be interpreted as a Lissajous plot of the solution), see
\figref{fig_in6}b. 

The first term in \eqref{in11} is a transient one, which decreases
exponentially fast. In fact, it is of order $\eps$ as soon as
$\tau\geqs\tau\sub1(\eps) = \eps\abs{\ln\eps}$. Since $\lim_{\eps\to
0}\tau\sub1(\eps) = 0$, we may write 
\begin{equation}
\label{in12}
\lim_{\eps\to 0} x(\tau;\eps) = \lambda(\tau) 
\qquad \text{for $\tau>0$}.
\end{equation} 
The state of the system is thus determined entirely by the slow variable
$\lambda$. According to the discussion of Subsection \ref{ssec_inhyst}, we
are in a situation without hysteresis, since the adiabatic and asymptotic
limit commute. Indeed, the physically meaningful procedure is to take the
asymptotic limit first: we find that trajectories converge to the periodic
solution $\bar{x}(\tau,\eps) = (\sin\tau - \eps\cos\tau)/(1+\eps^2)$. Then
we see that $\bar{x}(\tau,\eps)$ tends to $\lambda(\tau)$ in the adiabatic
limit $\eps\to 0$. On the other hand, taking the adiabatic limit directly in
\eqref{in10} yields the correct result $x = \lambda(\tau)$.

The fact that all orbits are attracted by a periodic one can also be seen on
the Poincar\'e map (taken at $\tau = 0 \equiv 2\pi$), which reads
\begin{equation}
\label{in13}
T: x \mapsto \biggbrak{x + \frac{\eps}{1+\eps^2}} \e^{-2\pi/\eps} -
\frac{\eps}{1+\eps^2},
\end{equation}
and admits a stable fixed point at $\fix{x} = -\eps/(1+\eps^2)$. 
Let us finally point out that the fact that the periodic solution
$\bar{x}(\tau;\eps)$ admits a convergent series in $\eps$ is rather
exceptional, in general we will only be able to obtain asymptotic series. 
\end{example}

\begin{figure}
 \centerline{ 
 \psfig{figure=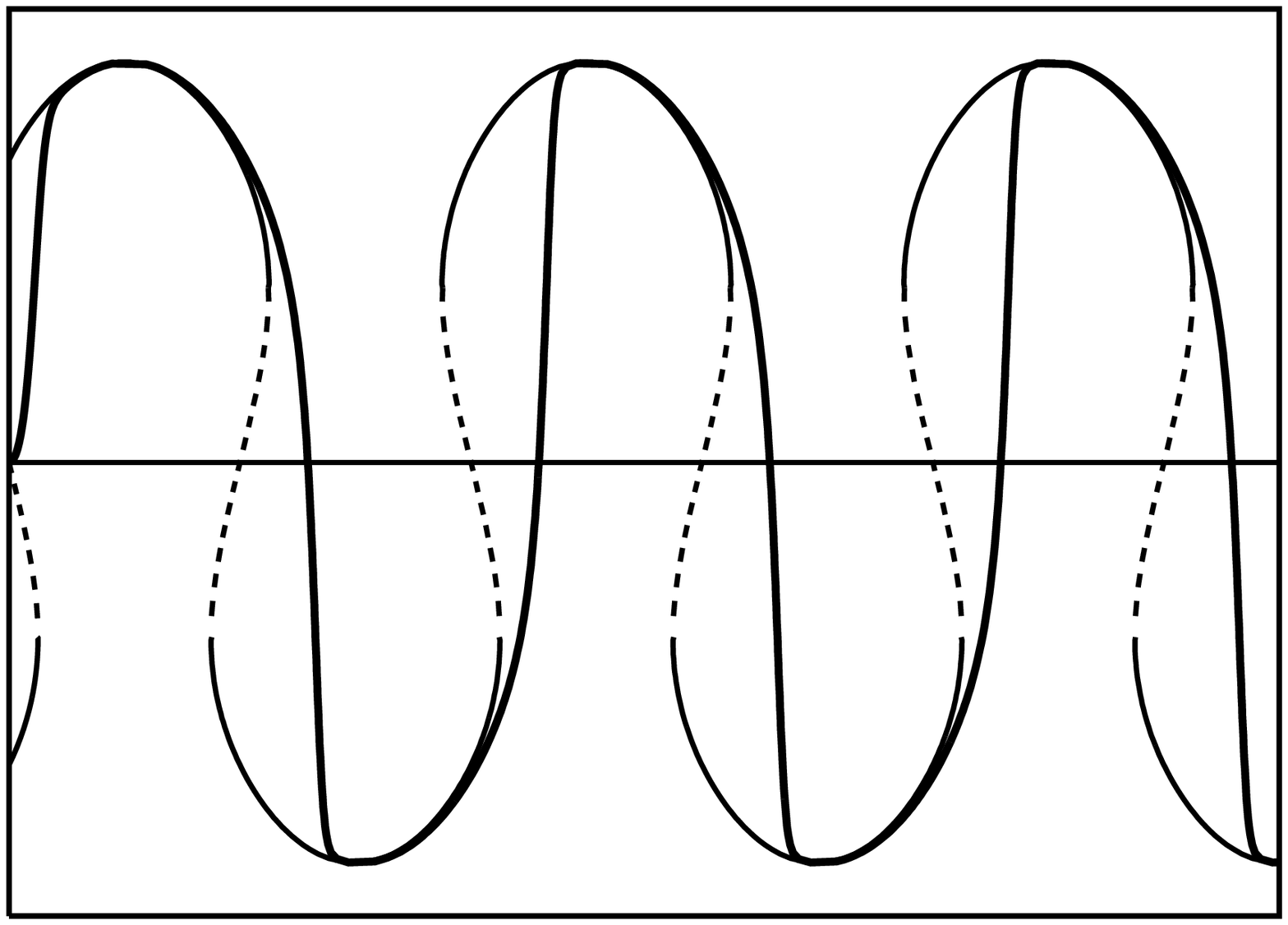,height=50mm,clip=t}
 \hspace{10mm}
 \psfig{figure=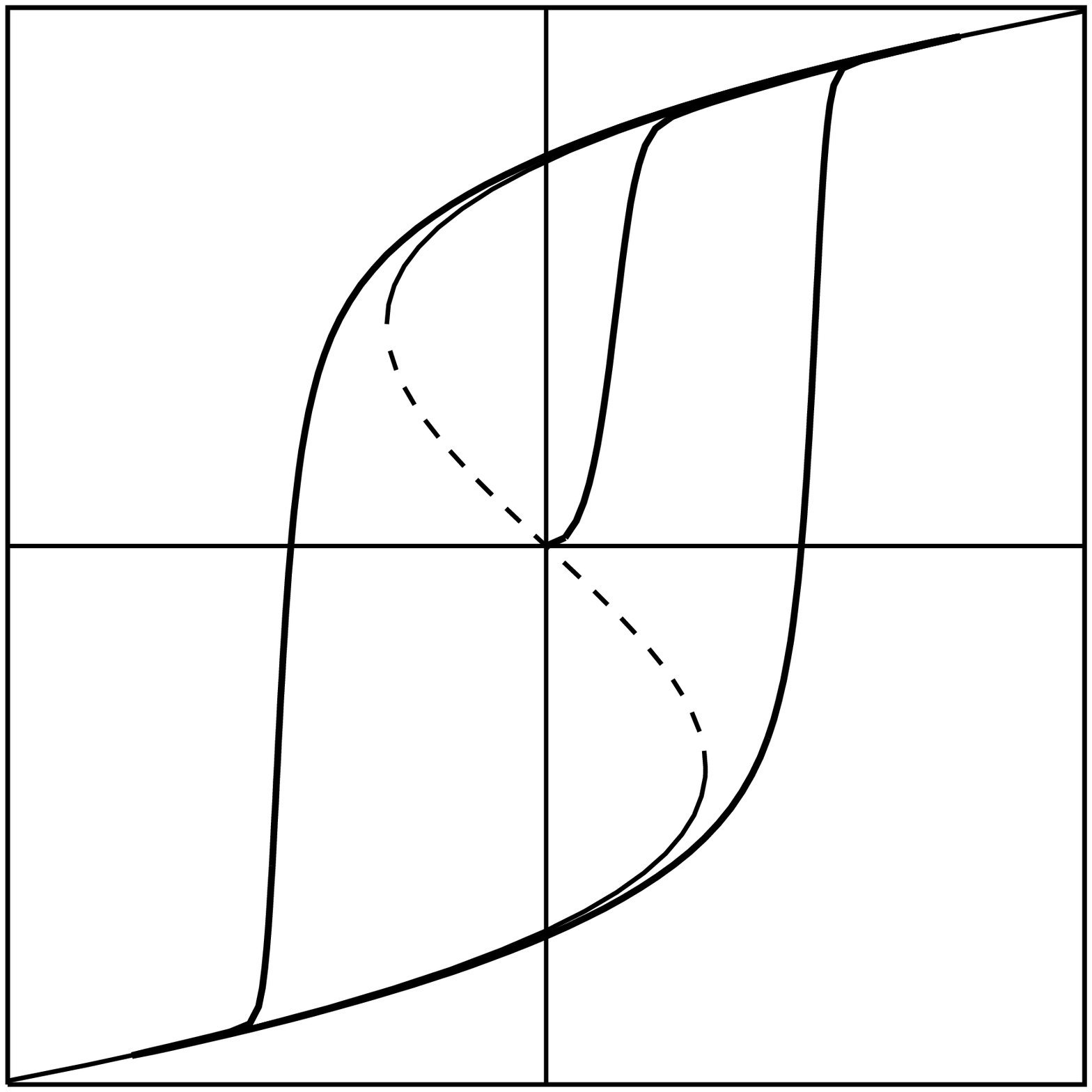,height=50mm,clip=t}
 }
 \figtext{
 	\writefig	0.5	5.2	a
	\writefig	8.6	5.2	b
 	\writefig	0.9	5.2	$x$
	\writefig	11.6	5.2	$x$
 	\writefig	7.5	3.1	$\tau$
	\writefig	13.6	3.1	$\lambda$
	\writefig	12.95	5.0	$\fix{x}_+(\lambda)$
	\writefig	9.1	0.9	$\fix{x}_-(\lambda)$
 }
 \caption[Solutions of the equation $\eps\dot{x} = x-x^3+\sin\tau$]
 {Solutions of equation \eqref{in14}, (a) in the $(\tau,x)$--plane
 and (b) in the $(\lambda,x)$--plane. Thin full lines indicate stable
 equilibria of the static system, dashed lines indicate unstable equilibria.
 These curves are solutions of the equation $x-x^3+\lambda = x-x^3+\sin\tau
 = 0$. The first representation (a) is useful to draw the vector field. One
 easily understands that the solution follows stable branches until the next
 saddle--node bifurcation, and then moves rapidly to the other branch. We
 obtain a periodic solution with alternating slow and fast motions, called a
 \defwd{relaxation oscillation}. When this solution is wrapped to the
 $(\lambda,x)$--plane, we obtain a familiar--looking hysteresis cycle. In
 the limit $\eps\to 0$, this cycle approaches a curve delimited by the
 equilibrium branches $\fix{x}_\pm(\lambda)$ and two verticals.}
\label{fig_in8}
\end{figure}

\newpage
\begin{example}
\label{ex_in2}
The equation
\begin{equation}
\label{in14}
\eps\dot{x} = x - x^3 + \lambda(\tau), 
\qquad \lambda(\tau) = \sin\tau,
\end{equation}
describes the overdamped motion of a particle in a
Ginzburg--Landau\index{Ginzburg--Landau potential} type double--well
potential $\Phi(x) = -\frac12x^2 + \frac14x^4$, with an external field
$\lambda$ (\figref{fig_in7}). This is the most common example for
hysteresis in ODE found in textbooks \cite{MR,MKKR}. 

\begin{figure}
 \centerline{\psfig{figure=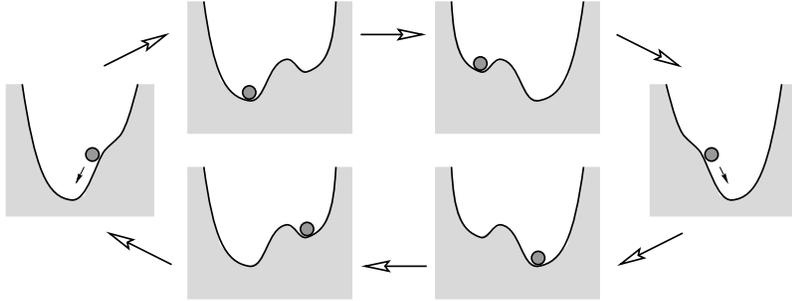,height=40mm,clip=t}}
 \figtext{
 }
 \caption[Potential associated with the equation $\eps\dot{x} =
 x-x^3+\sin\tau$]
 {Equation \eqref{in14} can be interpreted as describing the
 overdamped motion of a particle in a slowly varying potential of the form
 shown here. When $\lambda<-\crit{\lambda}$, the particle joins the
 equilibrium $\fix{x}_-(\lambda)$. For
 $-\crit{\lambda}<\lambda<\crit{\lambda}$, a new minimum $\fix{x}_+(\lambda)$
 has appeared, but the particle still remains in the left well. Only at
 $\lambda=\crit{\lambda}$, when the left equilibrium disappears, will the
 particle join $\fix{x}_+(\lambda)$, which it follows as long as
 $\lambda>-\crit{\lambda}$. For intermediate values of $\lambda$, the
 position of the particle depends not only on $\lambda(\tau)$, but also on
 $\dot{\lambda}(\tau)$: the system displays hysteresis.}
\label{fig_in7}
\end{figure}

Taking the adiabatic limit $\eps\to 0$ in \eqref{in14}, we obtain the
algebraic equation of a cubic $\lambda = -x + x^3$, admitting stationary
points $\pm(\crit{x},-\crit{\lambda})$, where $\crit{x} = 1/\sqrt{3}$ and
$\crit{\lambda} = 2/3\sqrt{3}$. When $\abs{\lambda}>\crit{\lambda}$, there is
a single solution $\fix{x}(\lambda)$, which corresponds to a stable
equilibrium of the static system. But when $\abs{\lambda}<\crit{\lambda}$,
there are {\em three} equilibrium curves (two stable and one unstable), and
it is not clear from this analysis which one the trajectory will follow. Let
us denote by $\fix{x}_+(\lambda)$ the upper stable equilibrium, and
$\fix{x}_-(\lambda)$ the lower one. 

Despite its simplicity, equation \eqref{in14} admits no exact solution. But
the qualitative behaviour of orbits can be easily understood by drawing the
vector field in the $(\tau,x)$--plane (\figref{fig_in8}a). Starting for
instance at $(\tau,x) = (0,0)$, the orbit will be attracted by the upper
branch $\fix{x}_+$, and follow it until it disappears when $\lambda$
becomes smaller than $-\crit{\lambda}$. If $\eps$ is small enough, the
trajectory will quickly reach the lower branch $\fix{x}_-$, and follow it
until $\lambda$ becomes larger than $\crit{\lambda}$ again. This behaviour
will repeat itself periodically, and it can be checked (using only
geometric properties of the Poincar\'e map) that the trajectory is
attracted by a periodic solution $\bar{x}(\tau;\eps)$. We thus obtain an
asymptotic cycle, characterized by alternating phases with slow and fast
motion. Such a solution is called a \defwd{relaxation
oscillation}\index{relaxation oscillation}.

When wrapping this solution to the $(\lambda,x)$--plane, we obtain that 
\begin{equation}
\label{in15}
\lim_{\eps\to 0} \bar{x}(\tau;\eps) = 
\begin{cases}
\fix{x}_+(\lambda) & \text{if $\lambda>\crit{\lambda}$ or
$\lambda>-\crit{\lambda}$ and $\dot{\lambda}<0$,} \\
\fix{x}_-(\lambda) & \text{if $\lambda<-\crit{\lambda}$ or
$\lambda<\crit{\lambda}$ and $\dot{\lambda}>0$.}
\end{cases}
\end{equation}
This solution displays the most familiar type of hysteresis. When
$\abs{\lambda}<\crit{\lambda}$, the asymptotic state (in the adiabatic
limit) depends not only on $\lambda$, but also on its
derivative.\footnote{If $\lambda(\tau)$ is a more complicated function than
$\sin\tau$, admitting several different maxima and minima, it may require
more information than $\lambda(\tau)$ and $\dot{\lambda}(\tau)$ to compute
the asymptotic state at time $\tau$. In fact, this state will depend on the
velocity of the last passage of $\lambda(\tau)$ through
$\pm\crit{\lambda}$.}

The limiting hysteresis cycle (\figref{fig_in8}b) has a well--defined area,
given by the geometric formula
\begin{equation}
\label{in16}
\cA(0) = 2\int_{-\crit{\lambda}}^{\crit{\lambda}} \fix{x}_+(\lambda)\dx
\lambda.
\end{equation}
It is clear from the vector field analysis that $\cA(\eps)$ increases with
$\eps$. In fact, it has been shown in \cite{Jung} that 
\begin{equation}
\label{in17}
\cA(\eps) \sord \cA(0) + \eps^{2/3}.
\end{equation}
We will show in Chapter 4 that this exponent $2/3$ can be computed
in a very simple way, using only local properties of the bifurcation points.
We point out that in this example, we have assumed the amplitude of
$\lambda(\tau)$ to be larger than $\crit{\lambda}$, so that $x(\tau)$
necessarily changes sign. We will examine in Chapter 7 what happens
when the amplitude approaches $\crit{\lambda}$. 
\end{example}

%\newpage

\begin{figure}
 \centerline{ 
 \psfig{figure=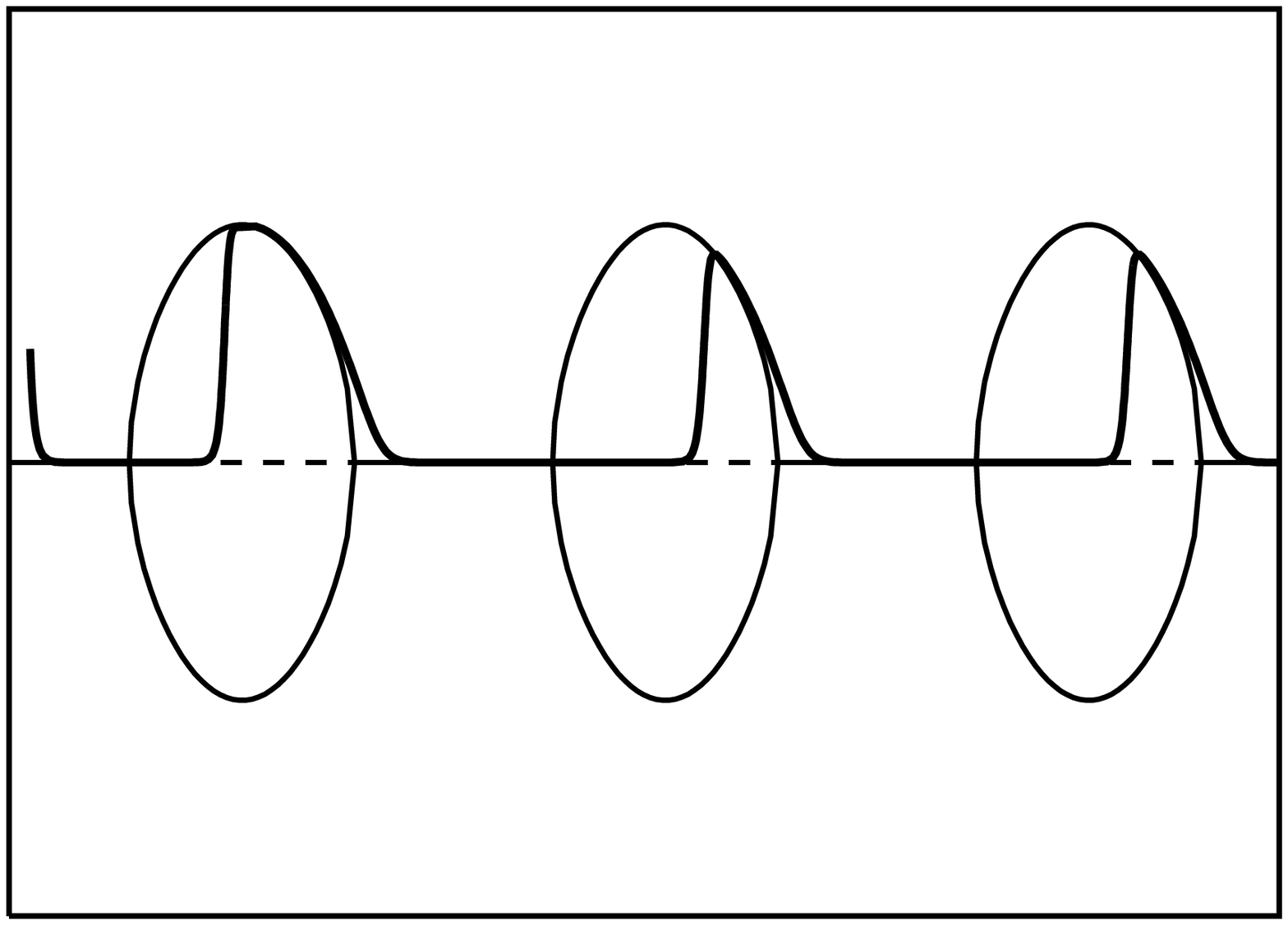,height=50mm,clip=t}
 \hspace{10mm}
 \psfig{figure=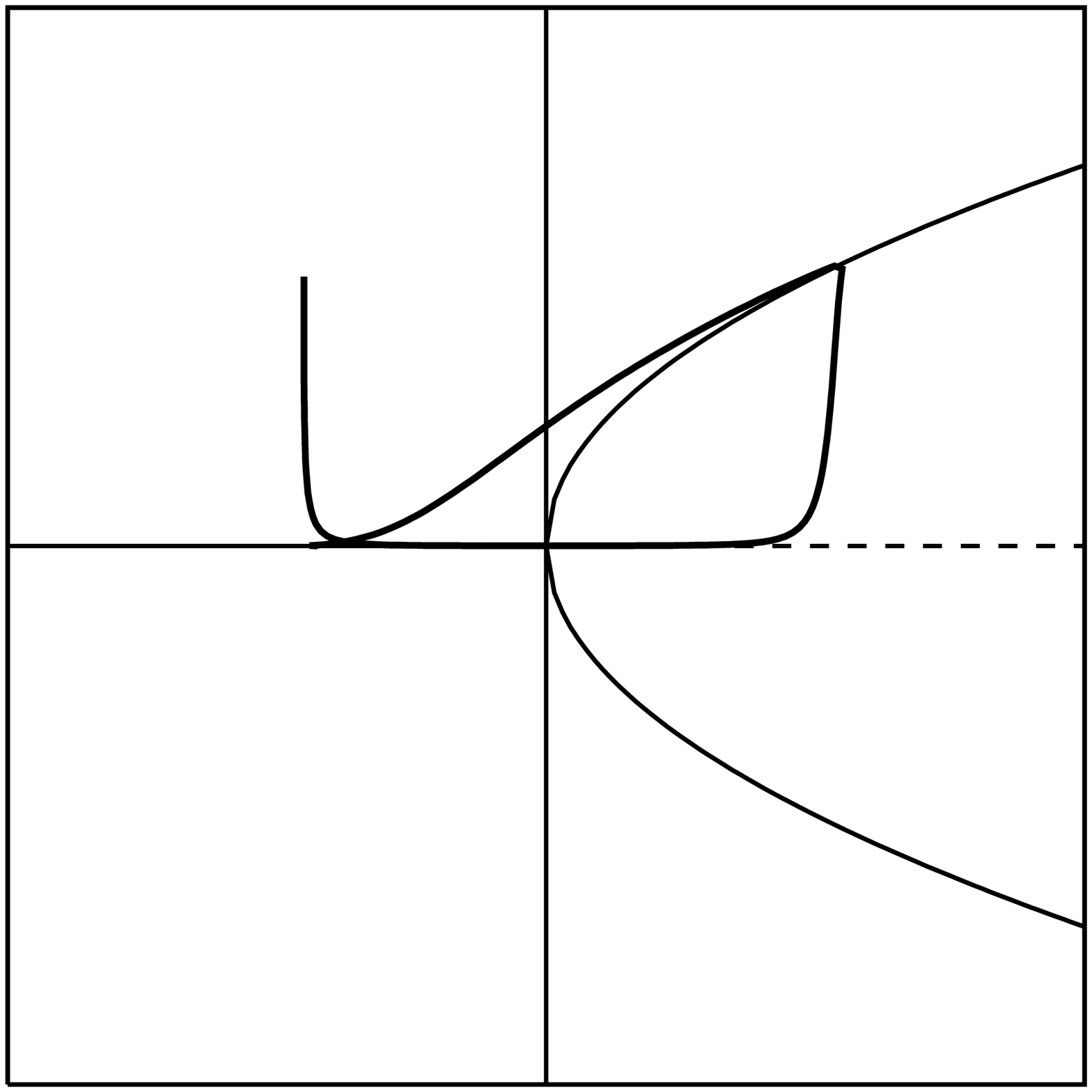,height=50mm,clip=t}
 }
 \figtext{
 	\writefig	0.5	5.2	a
	\writefig	8.6	5.2	b
 	\writefig	0.9	5.2	$x$
	\writefig	11.6	5.2	$x$
 	\writefig	7.5	3.1	$\tau$
	\writefig	13.6	3.1	$\lambda$
 }
 \caption[Solutions of the equation $\eps\dot{x} = \lambda(\tau)x-x^3$]
 {Solutions of equation \eqref{in18}, (a) in the $(\tau,x)$--plane
 and (b) in the $(\lambda,x)$--plane, but for the function $\lambda(\tau) =
 \sin\tau + 0.1$. Thin full lines indicate stable
 equilibria of the static system, dashed lines indicate unstable equilibria.
 These curves are solutions of the equation $\lambda x-x^3 
 = 0$. When the origin is stable, solutions reach it after a short time.
 They follow the origin for some macroscopic time after it has become
 unstable, a phenomenon known as \defwd{bifurcation delay}. If the solution
 finally jumps on another equilibrium, we obtain a hysteresis cycle.}
\label{fig_in10}
\end{figure}

\begin{example}
\label{ex_in3}
The equation
\begin{equation}
\label{in18}
\eps\dot{x} = \lambda(\tau)x - x^3, 
\qquad \lambda(\tau) = \sin\tau,
\end{equation}
describes the overdamped motion in a potential $\Phi(x,\lambda) =
-\frac12\lambda x^2 + \frac14x^4$. In the Ginzburg--Landau analogy, the
parameter $\lambda$ controls the temperature. The potential has a single
well at the origin if $\lambda<0$ ($T > T_{\math c}$, $T_{\math c}$ being
the critical temperature of a phase transition), and a double well if
$\lambda > 0$ ($T < T_{\math c}$). In the language of dynamical systems, we
have a pitchfork bifurcation at $\lambda = 0$. For positive $\lambda$, the
adiabatic system has to choose between two stable equilibria
$\pm\sqrt{\lambda}$ and the unstable origin.

Equation \eqref{in18} admits the explicit solution
\begin{equation}
\label{in19}
x(\tau) = \frac{x(\tau\sub0) \e^{\alpha(\tau)/\eps}}{\sqrt{1 +
\dss\frac{2}{\eps}x(\tau\sub0)^2 \int_{\tau\sub0}^{\tau}
\e^{2\alpha(s)/\eps} \dx s}}, 
\qquad
\alpha(\tau) \defby \int_{\tau\sub0}^{\tau} \lambda(s) \dx s 
= \cos\tau\sub0 - \cos\tau.
\end{equation}
It is not straightforward to analyse this solution analytically. Let us
consider the special case $\tau\sub0 = -\pi/2$, $\x(\tau\sub0) = 1$. Then
\begin{equation}
\label{in20}
x(\tau) = \frac{\e^{-\cos\tau/\eps}}{\sqrt{1 +
\dss\frac{2}{\eps} \int_{-\pi/2}^{\tau}
\e^{-2\cos s/\eps} \dx s}}% 
%= 
%\frac{1}{\sqrt{\e^{2\cos\tau/\eps} +
%\dss\frac{2}{\eps} \int_{-\pi/2}^{\tau}
%\e^{2(\cos\tau - \cos s)/\eps} \dx s}}
.
\end{equation}
For $-\pi/2 < \tau < \pi/2$, $-\cos\tau$ is negative, and the behaviour is
governed by the numerator $\e^{-\cos\tau/\eps}$, which is exponentially
small. Thus, the solution remains exponentially close to the origin until
$\tau = \pi/2$. For negative $\tau$, this is not surprising, since the
origin is stable. Although the origin becomes unstable at $\tau=0$, the
trajectory still remains close to it until $\tau=\pi/2$. This is a
simple example of \defwd{bifurcation delay}\index{bifurcation delay}: the
effective bifurcation takes place at $\tau=\pi/2$ rather than at $\tau=0$.
In the Ginzburg--Landau analogy, this phenomenon may be interpreted as
\defwd{metastability}\index{metastability}.

\begin{figure}
 \centerline{\psfig{figure=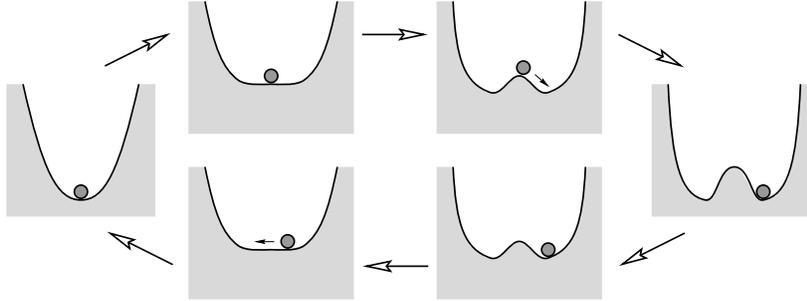,height=40mm,clip=t}}
 \figtext{
 }
 \caption[Potential associated with the equation $\eps\dot{x} = \sin(\tau)
 x - x^3$]
 {The potential corresponding to equation \eqref{in18}. For
 negative $\lambda$, the particle joins the single well at the origin. When
 this equilibrium becomes unstable, and two new wells are formed, the
 particle does nor react immediately to the bifurcation: it remains for some
 time in unstable equilibrium near the origin. This situation is called
 \defwd{delayed bifurcation}\index{delayed bifurcation}. Finally, the
 particle chooses a potential minimum, and follows it until the minima
 merge to form a single well again. This system displays hysteresis.}
\label{fig_in9}
\end{figure}

When $\tau>\pi/2$, the solution leaves the origin, and, in fact, settles
near the equilibrium position at $x = \sqrt{\sin\tau}$ until $\tau=\pi$,
when this branch merges with the origin again. One can show (using for
instance the saddle point method to estimate the integral in \eqref{in20})
that $x(\pi) = \Order{\eps^{1/4}}$. If we plot this solution in the
$(\lambda,x)$--plane, we find that the bifurcation delay leads to
\defwd{hysteresis}, since the trajectory always follows a stable branch for
decreasing $\lambda$, but sometimes follows an unstable one for increasing
$\lambda$. 

In fact, the solution analysed here is still a transient one. During the
next cycle of $\lambda$, the bifurcation delay is so large that the
trajectory ends up by always following the origin. But it is sufficient to
add an offset to $\lambda(\tau)$, of the form $\lambda(\tau) = \sin\tau +
c$, to obtain an asymptotic hysteresis cycle as in \figref{fig_in10}b. We
will show that its area scales as
\begin{equation}
\label{in21}
\cA(\eps) \sord \cA(0) + \eps^{3/4}.
\end{equation}
\end{example}

Considering the one--dimensional equations studied in these three examples,
we observe that
\begin{itemiz}
\item	solutions of a periodically forced system are always attracted by
	periodic ones;
\item	without bifurcations, the periodic solution encloses an area of
	order $\eps$, and does not display hysteresis in the adiabatic
	limit;
\item	when bifurcations are present, the periodic solution displays
	hysteresis, and encloses an area which follows a scaling relation of
	the form $\cA(\eps) \sord \cA(0) + \eps^{\mu}$, where $\mu$ is a
	nontrivial, fractional exponent. 
\end{itemiz}
One of the goals of this work will be to find out if these properties
remain valid for more general equations. We will see that asymptotic
solutions are not necessarily periodic. However, if such a periodic
solution exists, its area will usually follow a scaling law of the above
mentioned form, with an exponent $\mu$ which can be computed in a
relatively simple way.

%%%%%%%%%%%%%%%%%%%%%%%%%%%%%%%%%%%%%%%%%%%%%%%%%%%%%%%%%%%%%%%%%%%%%%%%%%%%%%%%%

%\newpage
\section{About this Thesis}
\label{sec_inabout}

%%%%%%%%%%%%%%%%%%%%%%%%%%%%%%%%%%%%%%%%%%%%%%%%%%%%%%%%%%%%%%%%%%%%%%%%%%%%%%%%%

%%%%%%%%%%%%%%%%%%%%%%%%%%%%%%%%%%%%%%%%%%%%%%%%%%%%%%%%%%%%%%%%%%%%%%%%%%%%%%%%%

\subsection{Objectives}
\label{ssec_inobj}

%%%%%%%%%%%%%%%%%%%%%%%%%%%%%%%%%%%%%%%%%%%%%%%%%%%%%%%%%%%%%%%%%%%%%%%%%%%%%%%%%

We pursue two major objectives in this work:
\begin{enum}
\item	Establish a coherent mathematical framework in order to deal with
	adiabatic systems of the form \eqref{in3}. In particular, we would
	like to understand the relation between an adiabatic system, and the
	corresponding family of autonomous equations. We are also interested
	in developing some practical tools allowing to establish existence
	of periodic orbits and hysteresis cycles, and to determine their
	scaling behaviour as a function of the adiabatic parameter.
	
\item	Apply these methods to some concrete examples. This should allow to
	check their efficiency to deal with a given equation, and to detect
	aspects of the theory which need further development. Since many
	authors, after spending much effort to derive equations describing
	magnetic hysteresis, analyse them by numerical simulations, we would
	like to show how the theory of \DynSys\ can be used to obtain
	valuable information on such equations with relatively small effort.	
\end{enum}
As we discussed in Subsection \ref{ssec_inhist}, much work has already been
done on adiabatic systems, in particular on bifurcation delay. We feel,
however, that this work is worth extending in two directions. Firstly,
results obtained by mathematicians are often formulated in a rather abstract
language, which is not easily accessible to the average (even theoretical)
physicist. Thus it is certainly useful to translate them into a language
facilitating their application to concrete problems. Secondly, several
aspects of the fundamental theory still need to be clarified. For instance,
hysteresis itself, and the associated scaling behaviour, have almost not
been studied by mathematicians. We also discovered, when analysing
particular examples, that several basic concepts still needed to be
developed, for instance adiabatic manifolds.

We have chosen two types of applications. The first one, which we call
nonlinear oscillators, concerns various situations where a damped particle
is placed in a slowly varying force field. Such low-dimensional \DynSys\ are
interesting for several reasons: we have some physical intuition for their
behaviour; they are sufficiently simple to be analysed in great detail, so
that we have a better chance to understand fundamental mechanisms of
hysteresis; still, some of these systems are known to exhibit chaotic motion
when forced periodically, and it is important to understand what happens
when this forcing becomes adiabatic. 

As a second application, we will consider a few models of magnetic
hysteresis. This program appears to be much more ambitious, since magnets
are so complicated systems that it is not clear at all whether they may be
modeled by finite dimensional equations. We think, however, that such an
attempt is justified by the mere fact that it will reveal both the power and
limits of such a kind of modeling. It may give some hints as to what
characteristics a realistic model should include, and in what directions the
theory should be extended in order to give more reliable predictions.

%%%%%%%%%%%%%%%%%%%%%%%%%%%%%%%%%%%%%%%%%%%%%%%%%%%%%%%%%%%%%%%%%%%%%%%%%%%%%%%%%

\subsection{Philosophy}
\label{ssec_inphil}

%%%%%%%%%%%%%%%%%%%%%%%%%%%%%%%%%%%%%%%%%%%%%%%%%%%%%%%%%%%%%%%%%%%%%%%%%%%%%%%%%

In this work, we adopt the point of view of Mathematical Physics. This
implies that physicists may regard it as an unnecessarily pedantic way of
establishing evidences, while mathematicians may consider it as a pedestrian
approach to a problem, which might be described much more nicely using
non--standard analysis and Borel series.

To the former, we would like to point out that there exist numerous
examples of problems, for which it was considered as evident that their
solutions behave in some special way, until this evidence was proved wrong
by a serious analysis. The precise mathematical understanding of a problem
is always desirable when it reveals the power and limits of an empirical
approach. For the latter, we would like to underline that our work aims at
providing a method of practical use, allowing the physicist to obtain
useful information on a concrete adiabatic system, with a minimum of
technical tools.

There are different approaches to Mathematical Physics. One of them relies
on exact solutions. We believe that this approach is useful as far as it
provides very precise information on a particular model equation, which is
assumed to be generic. There are, however, two major drawbacks: Firstly,
differential equation which can be solved exactly are very scarce, so that
only very few model equations are likely to be analysable in this way.
Secondly, even when a system has been solved exactly, the interesting
features are not immediately apparent, and it may require a lot of hard
analysis to derive them. This approach does not favour the physical
intuition, and often yields incorrect interpretations.\footnote{A startling
example of such a misunderstanding is found in \cite{AC1}, who analyse an
equation linearized around an {\em unstable} equilibrium, which is never
reached by the solution.}

A good illustration of these difficulties is provided by Example
\ref{ex_in3}: this system is still relatively simple to solve, if one
knows about Bernouilli's equation\index{Bernouilli
equation}\index{equation!Bernouilli}. But it turns out that the important
phenomenon, namely bifurcation delay, can also be obtained in a much simpler
way, by studying the linearized equation $\eps\dot{x} = \lambda(\tau)x$. The
behaviour of solutions far from the origin can be analysed by different
methods, that do not depend on the detailed form of the nonlinear term,
which is necessary for the equation to be exactly solvable. We will show
that even the scaling law \eqref{in21} can be obtained using only a local
analysis around the bifurcation point.

We will thus prefer those methods which favour the physical intuition. To
analyse some complicated equation, one has to understand first which terms
are important, and which terms have a negligible influence. Then one starts
by solving the simplified equation containing only the important terms.
Perturbative methods are often well adapted so such a procedure.

But one has to be careful not to confuse perturbation and approximation. It
is very tempting (and often done) to assume that a solution can be written
as power series of some small parameter, to insert this series into the
equation, and to solve it for the first few orders. This procedure is often
dangerous, since these power series do usually {\em not}
converge.\footnote{For instance, the perturbative analysis of a Hopf
bifurcation in \cite{ME1,BER} fails to reveal the phenomenon of maximal
delay.} In fact, it is better to apply perturbation theory to the equation
than to its solution.

We will often proceed in two steps. Firstly, we will derive an iterative
scheme that allows to decrease the order of some remainder in the
equation, which prevents us from solving it. Secondly, we have to prove in
an independent way that the influence of this small remainder on the
solution can be {\em bounded}. Thus, if we write that a solution contains a
remainder $R=\Order{\eps}$, we mean something very precise: namely there
exist positive constants $c$ and $\ez$ such that $\abs{R}\leqs c\eps$ for
$0<\eps\leqs\ez$. These constants are independent of $\eps$ and could be
computed explicitly (although this computation may turn out to be quite
cumbersome). 

Once such a bound on the remainder is known, when applying the theory to a
concrete example one can forget about the proof of the second part, and use
the iterative scheme to determine the behaviour of the solution at leading
order in the small parameter. It appears that the bounds $c$ and $\ez$ are
often far too pessimistic, and that, at least for finite dimensional
systems, the asymptotic theory provides rather accurate information for
fairly ``large'' values of the ``small'' parameter.

%%%%%%%%%%%%%%%%%%%%%%%%%%%%%%%%%%%%%%%%%%%%%%%%%%%%%%%%%%%%%%%%%%%%%%%%%%%%%%%%%

\subsection{Reader's Guide}
\label{ssec_inguide}

%%%%%%%%%%%%%%%%%%%%%%%%%%%%%%%%%%%%%%%%%%%%%%%%%%%%%%%%%%%%%%%%%%%%%%%%%%%%%%%%%

There are different ways to write a Ph.D.\ dissertation. One can choose to
present a summary of the major results, or one can give a more detailed
account with background information on the subject and complete proofs. We
preferred the second possibility. We believe that such a detailed
presentation is justified for the kind of subject we have been working on,
which lies on the boundary of Mathematics and Physics, provided the
structure of the text is sufficiently apparent. 

Although the chapters of this dissertation are not self--contained, we tried
to write them as far as possible in an independent way. Thus it is certainly
not necessary, in order to understand the contents of a given chapter, to
have read all the preceding ones. 

Roughly speaking, Chapters 2 and 3 present some
aspects of mathematical and physical theory which are already well known.
Chapters 4 and 5 are dedicated to the abstract,
mathematical theory which we have developed to deal with adiabatic \DynSys.
Chapters 6 and 7 provide applications of this theory to
some concrete examples. Chapter 8 contains extensions of some
results to iterated maps.

\subsubsection*{Chapter 2: Mathematical Tools}

We present some important notions from the theory of \DynSys\ (as
equilibrium points, stability and \Liap\ functions, invariant manifolds,
bifurcations and normal forms), and some elements of analysis which are
used in the proofs (Banach spaces, Fr\'echet derivatives, asymptotic series
and differential equations). The notations we use do in general not differ
from standard ones. This chapter should be considered as a reference
chapter, and the reader who is already well acquainted with the theory, as
well as the reader who is not interested in detailed mathematics, can safely
skip it.

\subsubsection*{Chapter 3: Physical Models}

We describe some physical concepts on which rely the examples discussed in
Chapters 6 and 7. The damped motion of a particle in a
potential serves as paradigm for a wide class of \DynSys. We discuss the
most common models for ferromagnets at equilibrium and out of equilibrium,
and show how to derive a deterministic evolution equation from the governing
master equation. Finally, we briefly present some existing phenomenological
models of hysteresis.

\begin{figure}
 \centerline{\psfig{figure=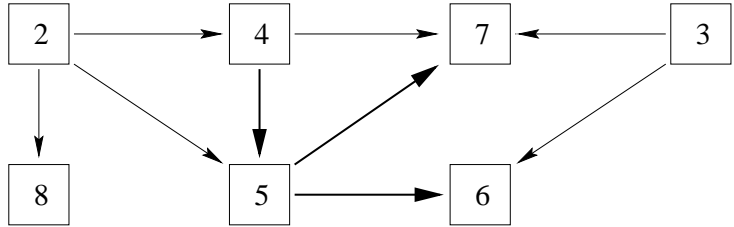,height=30mm,clip=t}}
 \vspace{2mm}
 \caption[Chapters of the thesis]
 {Logical organization of the chapters. Arrows indicate that a chapter
 relies on the contents of another one, in an essential way if the arrow is
 thick.}
\label{fig_in11}
\end{figure}

\subsubsection*{Chapter 4: One--Dimensional Systems}

We present a detailed mathematical framework to deal with one--dimensional
adiabatic equations of the form $\eps\dot{x} = f(x,\tau)$. We first discuss
the properties of adiabatic solutions, which are particular solutions
remaining close to non--bifurcating equilibria, and admitting asymptotic
series in the adiabatic parameter $\eps$. We then analyse in detail the
dynamics near bifurcation points, in particular the way how they scale with
$\eps$. We provide a simple geometrical method, based on the Newton polygon,
to determine the scaling exponents. Finally, we discuss some global aspects
of the flow, in particular how to determine periodic orbits, and the
$\eps$--dependence of hysteresis cycles.

\subsubsection*{Chapter 5: $n$--Dimensional Systems}

We extend results of the previous chapter to $n$--dimensional equations. The
discussion of adiabatic solutions is quite similar to the 1D case. We then
examine the linear equation $\eps\dot{y} = A(\tau)y$, which describes the
linearized motion around an adiabatic solution. This is a rather lengthy
task, but we show that the problem of diagonalizing such an equation can be
transformed into the problem of finding adiabatic solutions of an auxiliary
equation. This transformation allows to treat eigenvalue crossings and
bifurcations in a unified way. Next we develop some tools to deal with
nonlinear terms, in particular adiabatic manifolds and dynamic normal forms.
Finally, we examine some global properties of the flow.

\subsubsection*{Chapter 6: Nonlinear Oscillators}

We consider different examples involving the damped motion of a particle in
a slowly varying potential. The most important one is equivalent to a simple
physical system, namely a pendulum on a table rotating with slowly modulated
angular frequency. This system displays two important phenomena: a
bifurcation delay leading to hysteresis, and the possibility of a chaotic
motion, even for arbitrarily small adiabatic parameter. We use the methods
developed in Chapters 4 and 5 to compute an asymptotic
expression of the Poincar\'e map, which allows to delineate precisely the
parameter regions where hysteresis and chaos occur. The other two examples
discussed in this chapter illustrate the effect of eigenvalue crossings.

\subsubsection*{Chapter 7: Magnetic Hysteresis}

We discuss a few simple models of ferromagnets in a slowly varying magnetic
field. When the interaction between spins has infinite range (\ie in a
Curie--Weiss type model), the dynamics can be described in the thermodynamic
limit by a low--dimensional differential equation, of Ginzburg--Landau type.
We examine the phenomenon of dynamic phase transition for 1D spins, and the
effect of anisotropy on the dynamics of 2D spins. Finally, we explain why
models with short range interaction are much harder to analyse, and present
a few simple approximations.

\subsubsection*{Chapter 8: Iterated Maps}

We extend some of the previous results to adiabatic iterated maps. We start
by showing that some basic properties of adiabatic ODEs, such as existence
of adiabatic solutions and the behaviour of linear systems, can be extended
to maps depending on a slowly varying parameter. We conclude by presenting
some results on existence of adiabatic invariants for slow--fast maps, and
illustrate them on a few billiard problems.

%%%%%%%%%%%%%%%%%%%%%%%%%%%%%%%%%%%%%%%%%%%%%%%%%%%%%%%%%%%%%%%%%%%%%%%%%%%%%%%%%

\newpage
\section*{Acknowledgments}

%%%%%%%%%%%%%%%%%%%%%%%%%%%%%%%%%%%%%%%%%%%%%%%%%%%%%%%%%%%%%%%%%%%%%%%%%%%%%%%%%

This work would not have been possible without the help of many people,
to whom I would like to express my gratitude at this place. 

First of all, I thank my parents for offering me the possibility to study
Physics, and for their support during all these years. 

I am grateful to my thesis advisor, Professor Herv\'e Kunz, for accepting me
as a student and proposing me this interesting subject. I greatly benefited
from his broad scientific culture and honesty, and I thank him for taking
the time to discuss with me some of the problems I encountered, despite his
numerous other interests and busy academic life. 

I also thank Prof.\ J.--Ph.\ Ansermet, Prof.\ S.\ Aubry, Prof.\ B.\
Deveaud--Pl\'edran and Prof.\ O.E.\ Lanford III for accepting to be in the
thesis advisory board. 

My warmest thanks go to all members of the Institut de Physique Th\'eorique,
for the pleasant time I spent here. In particular, I thank Professors Ch.\
Gruber and Ph.A.\ Martin for offering me an interesting teaching activity;
Yvan Velenik, with whom I shared the office during so many years, for his
constant criticism which helped me to increase the standards of my
scientific research; Daniel Ueltschi and Claude--Alain Piguet for helping to
perpetuate the tradition of the weekly Ph.D.\ student's meeting; and, last
but not least, Christine Roethlisberger, the soul of the institute, for her
support in administrative problems, and her good temper. 

I am also grateful to Nilanjana Datta, Philippe Martin, Daniel Ueltschi and
Yvan Velenik for their critical reading of parts of this manuscript. 

I thank the Fonds National Suisse de la Recherche Scientifique for financial
support. 

My last thanks go to all my friends who helped me to remember, especially
during the last phase of my writing, that there still exists a world ``out
there''.

%%%%%%%%%%%%%%%%%%%%%%%%%%%%%%%%%%%%%%%%%%%%%%%%%%%%%%%%%%%%%%%%%%%%%%%%%%%%%%%%%

\end{document}